\documentclass[journal=jcisd8,manuscript=article]{achemso}

\usepackage[T1]{fontenc} % Use modern font encodings
\usepackage{chemformula} % Formula subscripts using \ch{}
\usepackage{siunitx}
\usepackage{subcaption}
\usepackage{booktabs}
\usepackage{caption}
\usepackage{threeparttable}
\usepackage{multirow}
\usepackage{nameref}
\usepackage{url}

\usepackage[]{todonotes}

\setkeys{acs}{maxauthors = 100}

\title{NNP/MM: Accelerating molecular dynamics simulations with machine learning potentials and molecular mechanics}

\author{Raimondas Galvelis}
\affiliation[AcelleraLabs]{Acellera Labs, C/ Doctor Trueta 183, 08005 Barcelona, Spain}
\email{r.galvelis@acellera.com}
\author{Alejandro Varela-Rial}
\affiliation[Acellera]{Acellera Ltd, Devonshire House 582, HA7 1JS, United Kingdom}
\alsoaffiliation[CSL]{Computational Science Laboratory, Universitat Pompeu Fabra, PRBB, C/ Doctor Aiguader 88, 08003 Barcelona, Spain}
\author{Stefan Doerr}
\affiliation[Acellera]{Acellera Ltd, Devonshire House 582,
HA7 1JS, United Kingdom}
\author{Roberto Fino}
\affiliation[AcelleraLabs]{Acellera Labs, C/ Doctor Trueta 183, 08005 Barcelona, Spain}
\author{Peter Eastman}
\affiliation[Stanford]{Department of Chemistry, Stanford University, 337 Campus Drive, Stanford, CA, 94305, USA}
\author{Thomas E. Markland}
\affiliation[Stanford]{Department of Chemistry, Stanford University, 337 Campus Drive, Stanford, CA, 94305, USA}
\author{John D. Chodera}
\affiliation[MSKCC]{Computational and Systems Biology Program, Sloan Kettering Institute, Memorial Sloan Kettering Cancer Center, New York, NY 10065, USA}
\author{Gianni De Fabritiis}
\affiliation[Acellera]{Acellera Ltd, Devonshire House 582,
HA7 1JS, United Kingdom}
\alsoaffiliation[CSL]{Computational Science Laboratory, Universitat Pompeu Fabra, PRBB, C/ Doctor Aiguader 88, 08003 Barcelona, Spain}
\alsoaffiliation[ICREA]{Instituci\'{o} Catalana de Recerca i Estudis Avan\c{c}ats (ICREA), Passeig Lluis Companys 23, 08010 Barcelona, Spain}
\email{g.defabritiis@gmail.com}

\begin{document}

\begin{tocentry}
\centering
\includegraphics[width=0.75\textwidth]{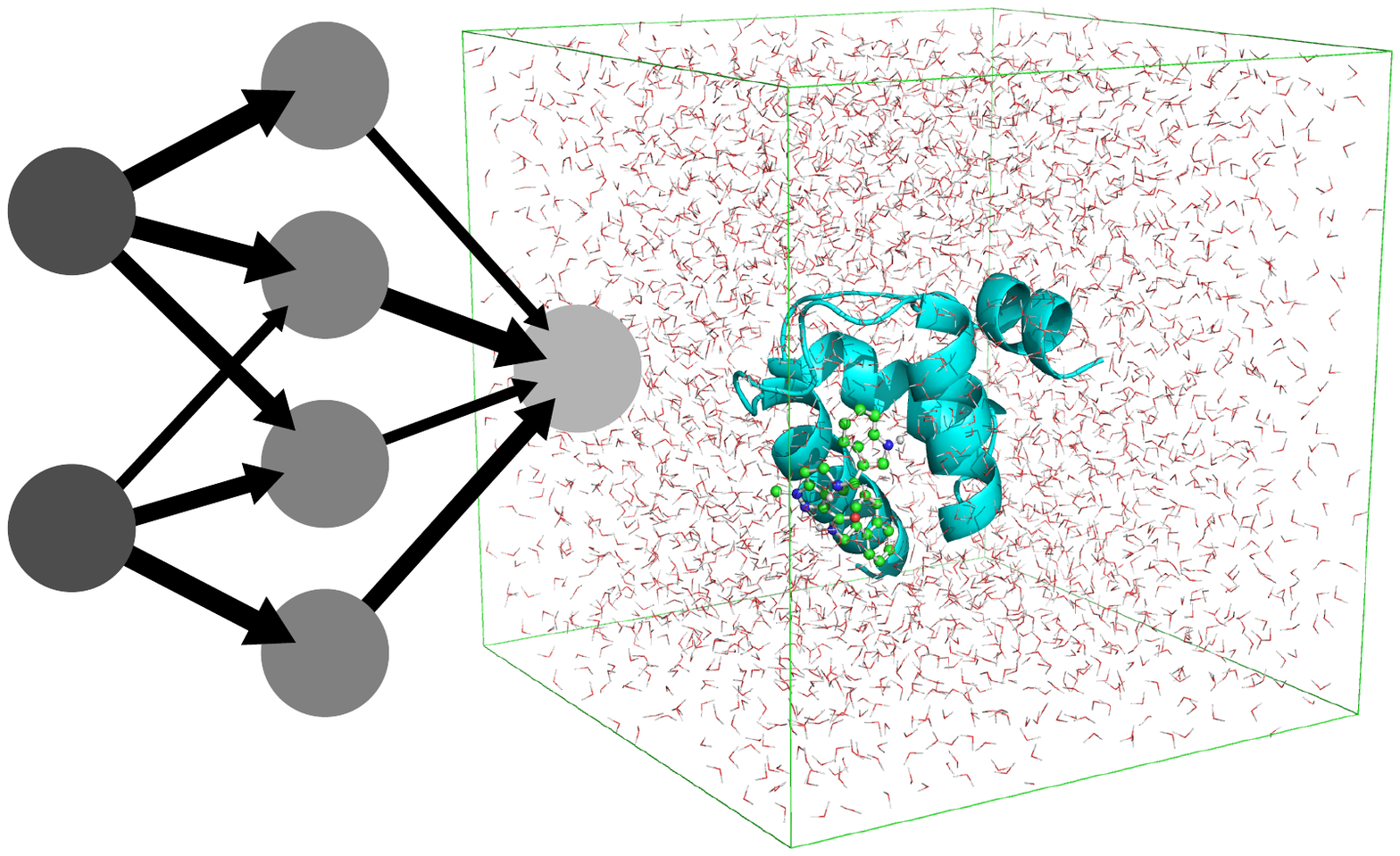}
\end{tocentry}

\begin{abstract}
Machine learning potentials have emerged as a means to enhance the accuracy of biomolecular simulations.
However, their application is constrained by the significant computational cost arising from the vast number of parameters compared to traditional molecular mechanics.
To tackle this issue, we introduce an optimized implementation of the hybrid method (NNP/MM), which combines neural network potentials (NNP) and molecular mechanics (MM).
This approach models a portion of the system, such as a small molecule, using NNP while employing MM for the remaining system to boost efficiency.
By conducting molecular dynamics (MD) simulations on various protein-ligand complexes and metadynamics (MTD) simulations on a ligand, we showcase the capabilities of our implementation of NNP/MM.
It has enabled us to increase the simulation speed by $\sim$5 times and achieve a combined sampling of \SI{1}{\micro\second} for each complex, marking the longest simulations ever reported for this class of simulation.
\end{abstract}

\section{Introduction}

%\todo[inline]{ACEMD}
In the past decade, molecular dynamics (MD) has transitioned from CPU execution to accelerators such as graphical processing units (GPUs).
Starting in 2006, CELLMD\cite{de2007performance} and subsequently ACEMD\cite{harvey2009acemd} began leveraging GPUs to enhance biomolecular simulations. Numerous other MD codes have either adapted (e.g., AMBER\cite{salomon2013overview}, GROMACS\cite{abraham2015gromacs}, NAMD\cite{phillips2020scalable}) or been initially designed to utilize GPUs (e.g., OpenMM\cite{eastman2010openmm}, HOOMD\cite{anderson2010hoomd}, TorchMD\cite{doerr2021torchmd}).
This innovation has improved the cost efficiency of MD simulations by two orders of magnitude\cite{stone2010gpu}.

%\todo[inline]{Advance of NNPs}
During the same timeframe, improvements in the accuracy of molecular mechanics (MM) and its force fields (FFs) have not advanced at a comparable pace as the simulation speed.
Widely adopted biomolecular FFs, such as AMBER\cite{cornell1995second, maier2015ff14sb}, CHARMM\cite{mackerell1998all, huang2013charmm36}, and others, offer parameters for proteins and common biomolecules.
However, obtaining accurate parameters for novel drug-like molecules remains a challenging task\cite{galvelis2019scalable}. The recent development of neural network potentials (NNPs) holds promise to address this issue\cite{noe2020machine}.
NNPs leverage the characteristic of neural networks (NNs) as \emph{universal approximators}, which means they can approximate any function with arbitrary precision relative to the training data.
In the context of molecular simulations, NNPs are designed to predict the energy and forces of quantum mechanics (QM) calculations\cite{behler2015constructing}.

Recently, numerous NNPs have been proposed (SchNet\cite{schutt2018schnet}, TensorMol\cite{yao2018tensormol}, AIMNet\cite{zubatyuk2019accurate}, 
PhysNet\cite{unke2019physnet}, DimeNet++\cite{klicpera2020fast}, OrbNet\cite{qiao2020orbnet}, PaiNN\cite{schutt2021equivariant}, SpookyNet\cite{unke2021spookynet}, NequIP\cite{batzner2021se}, OrbNet Denali\cite{christensen2021orbnet}, TorchMD-NET\cite{tholke2022equivariant}, MACE\cite{batatia2022mace},  etc).
One of the most used for organic molecules are ANI\cite{smith2017ani} and its derivatives\cite{smith2018less, smith2019approaching, stevenson2019schr, devereux2020extending} based on a modified Behler-Parrinello (BP) symmetry function\cite{behler2007generalized}.
For example, the benchmarks of ANI-2x on a set of biaryl fragment, typically found in drug molecules, shows better accuracy than the established general small molecule FFs (CGenFF\cite{vanommeslaeghe2010charmm, vanommeslaeghe2012automation}, GAFF\cite{wang2004development}, OPLS\cite{jorgensen2005potential}, and OpenFF\cite{mobley2018escaping}).
The mean absolute error for the entire potential energy profile and rotational barrier heights are \SI{0.5}{kcal/mol} and \SI{1.0}{kcal/mol}, respectively\cite{lahey2020benchmarking}, but it is orders of magnitude faster than its reference QM calculations at the DFT level ($\omega$B97X/6-31G*)\cite{devereux2020extending}.
 However, the BP-type NNPs have several limitations. First, the long-range interactions are not properly accounted for.
The NNPs only consider the chemical environment around each atom within a given cut-off distance (\SI{5.1}{\angstrom} for ANI-2x).
Second, a limited set of elements is supported (H, C, N, O, F, S, and Cl for ANI-2x).
Finally, only neutral molecules can be computed\cite{devereux2020extending}.

%\todo[inline]{Application of NNP/MM for biomolecule simulations}
Despite the current limitations, NNPs are already improving biomolecular simulations.
It has demonstrated that the accuracy for drug-like molecules is improved\cite{galvelis2019scalable} by reparameterizing dihedral angles with ANI-1x\cite{smith2018less}.
Alternatively, the hybrid method of NNP and MM (NNP/MM)\cite{lahey2020simulating} allows embedding NNP into a simulation.
The main idea of NNP/MM is similar to QM/MM\cite{warshel1976theoretical, lin2007qm, senn2009qm}: an important region of a system is modeled with a more accurate method, while a less accurate and computationally cheaper one is used for the rest of the system.

%\todo[inline]{Previous works}
Recently, \citeauthor{lahey2020simulating}\cite{lahey2020simulating} have demonstrated the first application of NNP/MM to protein-ligand complexes.
NNP/MM is used to refine binding poses and to compute the conformational free energies.
\citeauthor{rufa2020towards}\cite{rufa2020towards} have computed the binding free energies of the Tyk2 congeneric ligand benchmark series\cite{wang2015accurate} using alchemical free energy calculations.
Instead of using NNP/MM directly, a non-equilibrium switching scheme has been devised to correct the standard MM calculations to NNP/MM accuracy.
It reduces the errors from \SI{1.0}{kcal/mol} to \SI{0.5}{kcal/mol}.
\citeauthor{vant2020flexible}\cite{vant2020flexible} have used NNP/MM for the refinement of a protein-ligand complex from cryo-electron microscopy data.
The refinement with NNP/MM produces higher-quality models than QM/MM with the semi-empirical PM6 method at a lower computational cost.
\citeauthor{xu2021automatically}\cite{xu2021automatically} have trained a specialized NNP for zinc and used NNP/MM to simulate zinc-containing proteins.
The obtained results are in agreement with QM/MM calculations. 

%\todo[inline]{Current performance}
A critical limitation for the wider adoption of NNP/MM is the simulation speed.
Despite NNP being much faster than QM, it is still slower than MM.
For example, \citeauthor{lahey2020simulating}\cite{lahey2020simulating} and \citeauthor{vant2020flexible}\cite{vant2020flexible} have reported the simulations speed of \SI{3.4}{ns/day} and \SI{0.5}{ns/day}, respectively, on an NVIDIA TITAN Xp GPU.
Also, the longest reported simulation is just \SI{20}{ns}\cite{vant2020flexible}.

%\todo[inline]{Hypothesis}
In this work, we present an optimized implementation of NNP/MM in ACEMD\cite{harvey2009acemd} based on OpenMM\cite{eastman2010openmm} and PyTorch\cite{paszke2019pytorch}.
First, the method and relevant optimization strategies are introduced.
Second, the capability of software is demonstrated by performing metadynamics (MTD)\cite{barducci2011metadynamics} simulations of a fragment of erlotinib and molecular dynamics (MD) simulations of four protein-ligand complexes.
Finally, the installation and setup of simulations are shown.

\section{Methods}

%\todo[inline]{Potential energy}
In the NNP/MM approach, a system is partitioned into NNP and MM regions similarly to QM/MM\cite{warshel1976theoretical, lin2007qm, senn2009qm}.
The total potential energy ($V$) consists of three terms: 
\begin{equation}\label{eq:nnp-mm}
    V(\vec r) = V_\text{NNP}(\vec r_\text{NNP}) + V_\text{MM}(\vec r_\text{MM}) + V_\text{NNP-MM}(\vec r)
\end{equation}
where $V_\text{NNP}$ and $V_\text{MM}$ are the potential energies of the NNP and MM regions, respectively.
$V_\text{NNP-MM}$ is a coupling term; $\vec r$, $\vec r_\text{NNP}$, and $\vec r_\text{MM}$ are the atomic position of the entire system, NNP region, and MM region, respectively.

%/todo[inline]{NNP model}
It is required that the NNP potential ($\vec r_\text{NNP}$) is a function of the atomic position ($\vec r_\text{NNP}$) and atomic numbers ($\vec Z_\text{NNP}$) of the NNP region.
The total charge ($q_\text{NNP}$) can be included if necessary:
\begin{equation}
    V_\text{NNP}(\vec r_\text{NNP}) \equiv V_\text{NNP}(\vec Z_\text{NNP}, \vec r_\text{NNP}, q_\text{NNP})
\end{equation}
Also, it is required that $V_\text{NNP}$ is differentiable with respect to $\vec r_\text{NNP}$ to compute the atomic forces ($\vec F_\text{NNP}$):
\begin{equation}
    \vec F_\text{NNP} = - \nabla V_\text{NNP}
\end{equation}

%\todo[inline]{Embedding scheme}
In this work, we adapt the coupling term ($V_\text{NNP-MM}$) proposed by \citeauthor{lahey2020simulating}\cite{lahey2020simulating}:
\begin{equation}\label{eq:embeding}
    V_\text{NNP-MM}(\vec r) = \sum^{N_{\text{NNP}}}_i \sum^{N_{\text{MM}}}_j
        \left(
            V_C^{i,j} + V_{LJ}^{i,j}
        \right) 
\end{equation}
where $V_C^{i,j}=\frac{q_i q_j}{4 \pi \epsilon_0 r_{ij}}$ is the  Coulomb potential, $V_{LJ}^{i,j}=4 \epsilon_{ij}\left[\left( \frac{\sigma_{ij}}{r_{ij}} \right)^{12} - \left( \frac{\sigma_{ij}}{r_{ij}} \right)^6\right]$ is the Lennard-Jones potential and $N_{\text{NNP}}$ and $N_{\text{MM}}$ are the number of NNP and MM atoms, respectively; $q_i$ and $q_j$ are the atomic charges; $\epsilon_{ij}$ and $\sigma_{ij}$ are the Lennard-Jones parameters; $r_{ij}$ is the distance between the atoms; and $\epsilon_0$ is the vacuum permittivity (dielectric constant).
In the context of QM/MM, this is known as the \emph{mechanical embedding} scheme\cite{lin2007qm,senn2009qm}.

%\todo[inline]{Software stack}
NNP/MM is implemented in ACEMD\cite{harvey2009acemd} using several software components.
OpenMM\cite{eastman2017openmm}, a GPU-accelerated MD library, is used to compute MM terms and propagate the MD trajectory.
OpenMM-Torch\cite{openmm-torch}, an OpenMM plugin, is used to compute the NNP term.
It uses PyTorch\cite{paszke2019pytorch}, a machine learning framework for NN training and inference on GPUs, to load and execute the NNP on GPU.
TorchANI\cite{gao2020torchani} is used to create the PyTorch model of ANI-2x\cite{devereux2020extending}.
NNPOps\cite{nnpops}, a library of optimized CUDA kernels for NNP, is used to accelerate critical parts of the computations.
Future versions will integrate other NNPs available in TorchMD-NET\cite{tholke2022equivariant, torchmd-net}.

%\todo[inline]{Optimization}
We have optimized the performance of NNP/MM in three ways.
First, all the terms of NNP and MM are computed on a GPU.
Neither atomic positions nor atomic forces need to be transferred between the CPU and GPU, as is the case with the original implementation\cite{lahey2020simulating}.
Second, the featurizer of ANI has been implemented as a custom CUDA kernel and is available in the NNPOps library\cite{nnpops}.
The original featurizer in TorchANI is implemented using only standard PyTorch operations, which are an inefficient way of performing this calculation.
Third, the computation is parallelized over the NNs (ANI-2x has an ensemble of 8 NNs) and atoms taking advantage that the same molecule is computed repeatedly.
The original implementation in TorchANI computes the NNs sequentially.
The original TorchANI version is optimized for batch computing, i.e. many molecules are computed simultaneously, while for MD low-latency computing, i.e. one molecule is computed as fast as possible is necessary.
The weights and biases of the atomic NNs are replicated and batched in the same order as the atoms in a molecule, allowing a GPU to efficiently parallelize the calculation for a single molecule.
The implementation of the optimized NNs is available in the NNPOps library (\url{https://github.com/openmm/nnpops}).

\section{Results and Discussion}

\subsection{Simulations of a fragment}

\begin{figure}
    \centering
    \includegraphics[width=0.25\textwidth]{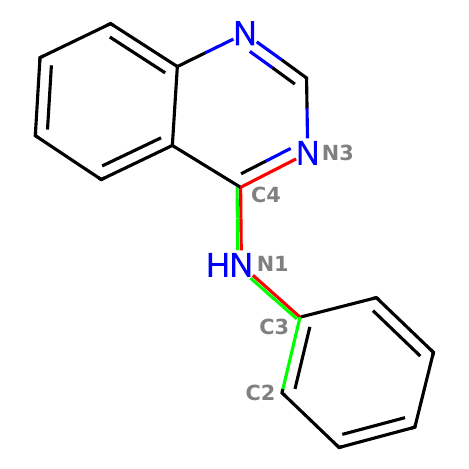}
    \caption{A fragment of erlotinib. The two dihedral angles used as the collective variable are shown in red and green.}
    \label{fig:fragment}
\end{figure}

%\todo[inline]{Fragment}
We use metadynamics\cite{barducci2011metadynamics}(MTD) to simulate a fragment (Figure~\ref{fig:fragment}) of erlotinib using two models: (1) the conventional MM with GAFF2\cite{wang2004development} parameters for the fragment; and (2) the NNP/MM where the fragment is modeled with ANI-2x\cite{devereux2020extending}.
\citeauthor{lahey2020simulating}\cite{lahey2020simulating} reported that the fragment has a notable discrepancy between the potential energy surfaces of CGenFF\cite{vanommeslaeghe2010charmm} and ANI-1ccx\cite{smith2019approaching}.
In this work, we expand the benchmark by computing the free energy surfaces.

%\todo[inline]{MTD}
We use the well-tempered MTD\cite{barducci2008well} with two dihedral angles (Figure~\ref{fig:fragment}) as collective variables.
The MTD simulations use the NVT ensemble (T~=~\SI{310}{K}), the time step is set to \SI{4.0}{fs} for the MM simulations, and to \SI{2.0}{fs} for the NNP/MM simulations because they are unstable with \SI{4.0}{fs}.
For MTD, PLUMED\cite{tribello2014plumed} is used.
More details are provided in the supplementary information.

The fragment was simulated for \SI{100}{ns} with each method.
This is sufficient to achieve extensive sampling in the collective variable space.
The time series of the dihedral angles (Figure~\ref{fig:fragment}) are available in the supplementary information (Figure~S1-S2).

\begin{figure*}
    \centering
    \begin{subfigure}{0.4\textwidth}
        \includegraphics[width=\textwidth]{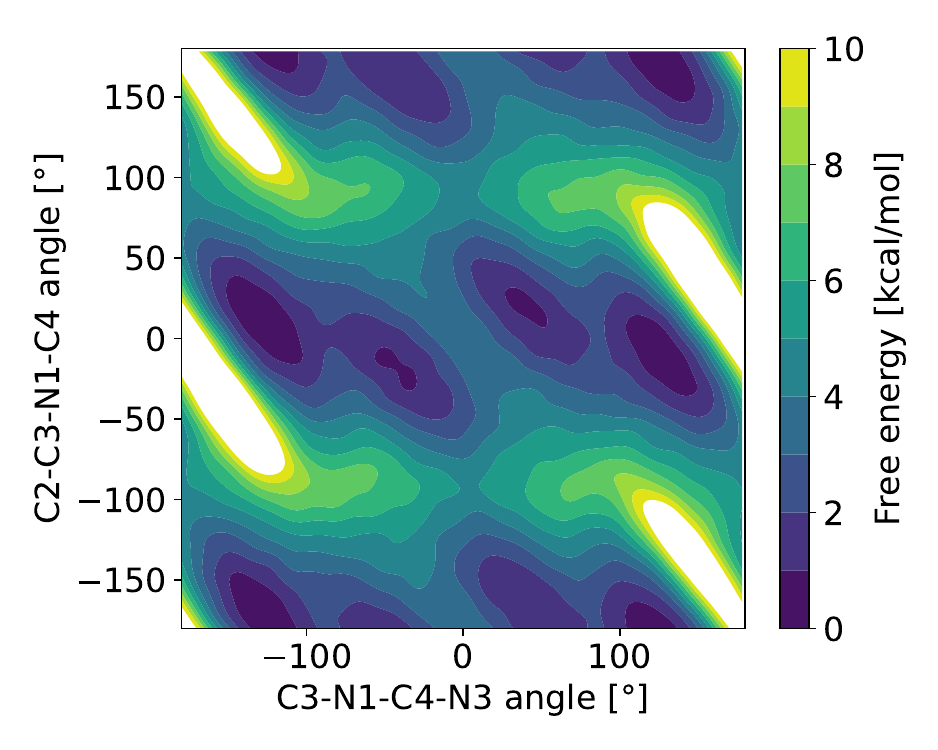}
        \caption{MM}
    \end{subfigure}
    \begin{subfigure}{0.4\textwidth}
        \includegraphics[width=\textwidth]{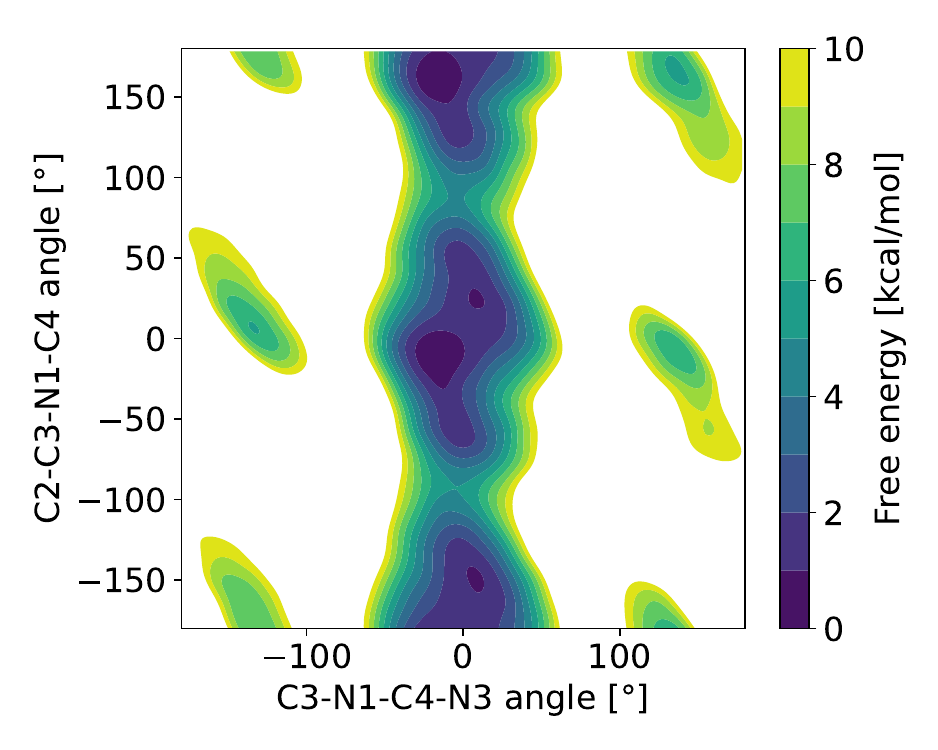}
        \caption{NNP/MM}
    \end{subfigure}
    \caption{Free energy surface of a fragment (Figure~\ref{fig:fragment}) of erlotinib computed with MTD using two models MM (a) and NNP/MM (b).}
    \label{fig:mtd}
\end{figure*}

%\todo[inline]{FES}
The obtained free energy surfaces (Figure~\ref{fig:mtd}) show a significant difference between the models.
The dominant conformer of the dihedral angle C3-N1-C4-N3 is predicted by GAFF2 and ANI-2x at $\sim$\SI{120}{\degree} and $\sim$\SI{0}{\degree}, respectively.
The fragment has two aromatic rings connected by a conjugated linker, so a planar conformation is expected to be energetically favorable.
This is consistent with the potential energy surfaces reported by \citeauthor{lahey2020simulating} (see Ref.~\citenum{lahey2020simulating}, Figure~3b).
Note, the fragment has been chosen for demonstration only and further analysis is beyond the scope of this work.

\subsection{Simulations of protein-ligand complexes}

\subsubsection{Protein-ligand complexes}

\begin{figure}
    \centering
    \begin{subfigure}{0.3\textwidth}
        \includegraphics[width=\textwidth, trim=4cm 6cm 4cm 6cm, clip=true]{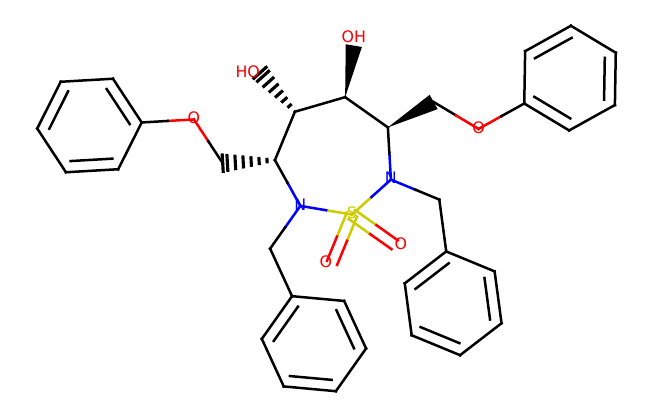}
        \caption{1AJV}
    \end{subfigure}
    \begin{subfigure}{0.3\textwidth}
        \includegraphics[width=\textwidth, trim=4cm 6cm 4cm 6cm, clip=true]{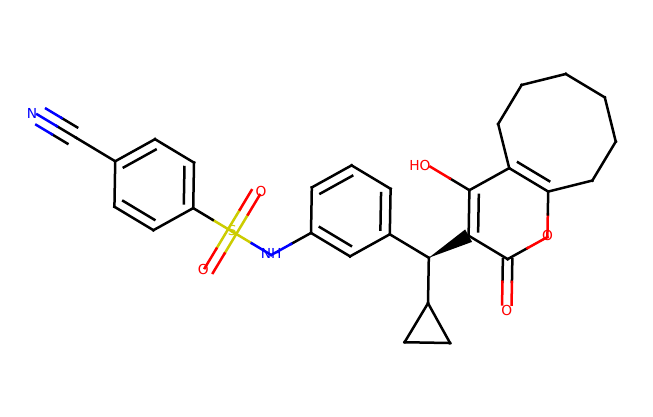}
        \caption{1HPO}
    \end{subfigure}
    \\
    \begin{subfigure}{0.3\textwidth}
        \includegraphics[width=\textwidth, trim=4cm 6cm 4cm 6cm, clip=true]{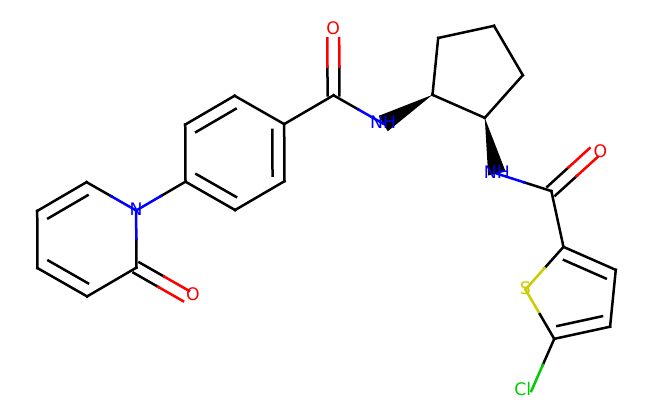}
        \caption{2P95}
    \end{subfigure}
    \begin{subfigure}{0.3\textwidth}
        \includegraphics[width=\textwidth, trim=4cm 6cm 4cm 6cm, clip=true]{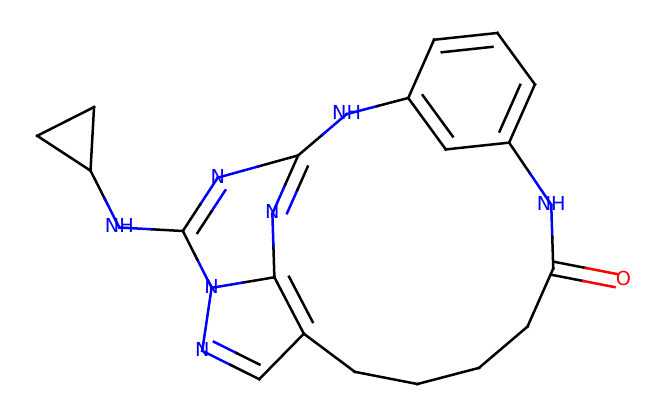}
        \caption{3BE9}
    \end{subfigure}
    \caption{Ligand structures (a-d) of the selected protein-ligand complexes.}
    \label{fig:ligands}
\end{figure}

%\todo[inline]{Selection criteria}
We have selected four protein-ligand complexes from PDBbind-2019\cite{wang2004pdbbind, liu2017forging} following these criteria.
First, the ligand contains only elements supported by ANI-2x (H, C, N, O, F, S, and Cl)\cite{devereux2020extending} and no charged functional groups (amine, carboxylate, etc).
Second, the ligand has less than one hundred atoms.
Third, the ligand has at least one rotatable bond, and the rotamers energies differ by >\SI{3}{kcal/mol} between GAFF2\cite{wang2004development} and ANI-2x\cite{devereux2020extending}.
We use the \emph{Parameterize} tool \cite{galvelis2019scalable} to detect the rotatable bond, scan the dihedral angles of rotatable bonds, and compute the relative rotamer energies.
The summary of the protein-ligand complexes is given in Table~\ref{tab:sys} and the ligand structures are shown in Figure~\ref{fig:ligands}. 
Additionally, the energy profiles of the dihedral angle scan of the ligands are available in the supplementary information (Figure~S2-S22).

\begin{table*}
\caption{Summary of protein-ligand complexes.}
\label{tab:sys}
\begin{threeparttable}
\begin{tabular}{c@{\hskip 0.05\textwidth}
                >{\centering}p{0.1\textwidth}
                >{\centering}p{0.1\textwidth}@{\hskip 0.05\textwidth}
                >{\centering}p{0.1\textwidth}
                >{\centering}p{0.1\textwidth}@{\hskip 0.05\textwidth}
                c}
\toprule
\multirow{2}{*}{System} & \multicolumn{2}{c}{Protein} & \multicolumn{2}{c}{Ligand} & \multirow{2}{0.1\textwidth}{\centering Total atoms} \\
 & atoms & residues & atoms & dihedrals & \\
\midrule
1AJV\cite{1ajv}      &  3125 &   198 &    75 &     5 & 38325 \\
1HPO\cite{1hpo}      &  3133 &   198 &    64 &     6 & 47712 \\
2P95\cite{2p95}      &  4398 &   286 &    50 &     7 & 52477 \\
3BE9\cite{3be9}      &  5451 &   328 &    48 &     2 & 60412 \\
\bottomrule
\end{tabular}
\end{threeparttable}
\end{table*}

%\todo[inline]{System preparation and simulation}
The protein-ligand complex preparation and equilibration have been carried out with HTMD\cite{doerr2016htmd}.
Each complex has been simulated with two different methods: MM, where the ligand is parameterized with GAFF2\cite{wang2004development} and NNP/MM, where the ligand is modeled with ANI-2x\cite{devereux2020extending}.
The protein, in both cases, uses AMBER ff14SB\cite{maier2015ff14sb} FF.
The MD simulations use the NVT ensemble (T~=~\SI{310}{K}), the time step is set to \SI{4.0}{fs} for the MM simulations, and to \SI{2.0}{fs} for the NNP/MM simulations because they are unstable with \SI{4.0}{fs}. 
For each combination of a complex and method, 10 independent simulations of \SI{100}{ns} are performed resulting in the combined sampling of \SI{1}{\micro\second}.
More details are provided in the supplementary information.

\subsubsection{Analysis of protein-ligand complexes}

All the proteins and ligands maintain their structures in the simulations with both methods (MM and NNP/MM).
The protein RMSD fluctuates in the range of \SIrange{1.8}{2.8}{\angstrom} and the residue RMSF have similar magnitudes when comparing the same protein with both methods.
The ligand RMSD fluctuates in the range of \SIrange{0.2}{1.7}{\angstrom}.
In the case of 1AJV and 2P95, there is no significant difference between MM and NNP/MM, but, in the case of 1HPO and 3BE9, the fluctuations are larger by \SI{\sim 0.3}{\angstrom} for NNP/MM.
The time series of protein RMSD, residue RMSF, and ligand RMSD are available in the supplementary information (Figure~S23-S34).
The difference of the ligand RMSD is expected because, as previous works\cite{galvelis2019scalable, lahey2020benchmarking, rufa2020towards} indicates, ANI-2x models the dihedral angles more accurately than GAFF.
Also, it is important to note that our simulations are 50 times longer than previously reported\cite{lahey2020simulating} and have not resulted in any non-physical conformation.

\begin{figure*}[tb]
    \centering
    \begin{subfigure}{0.4\textwidth}
        \vspace{-0.5cm}
        \includegraphics[width=\textwidth]{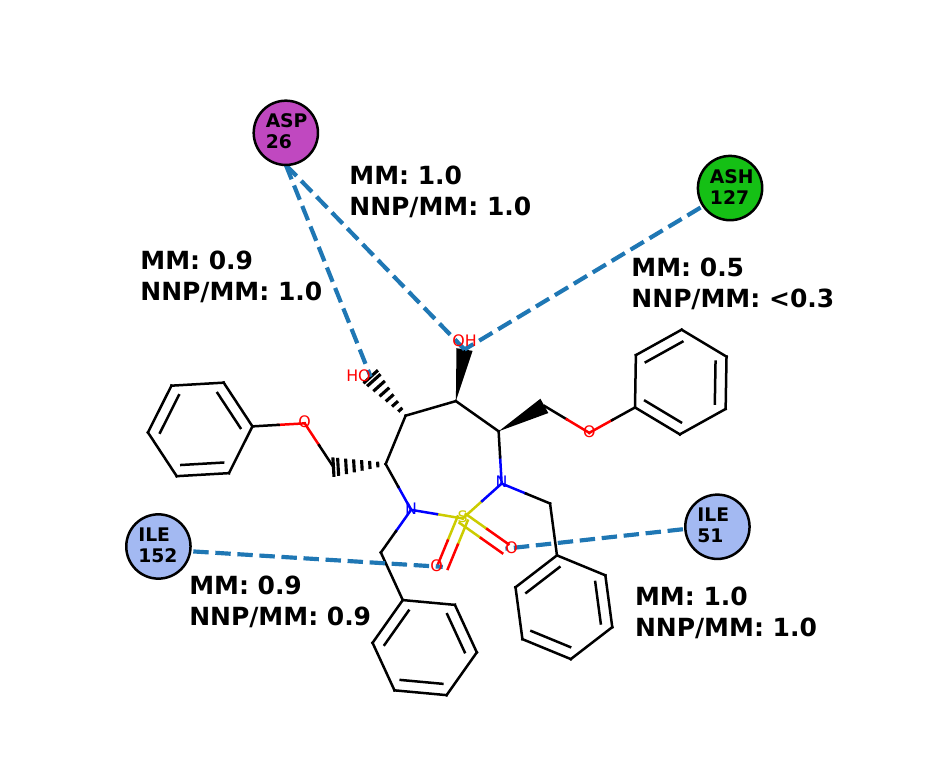}
        \vspace{-0.5cm}
        \caption{1AJV}
    \end{subfigure}
    \begin{subfigure}{0.4\textwidth}
        \vspace{-0.5cm}
        \includegraphics[width=\textwidth]{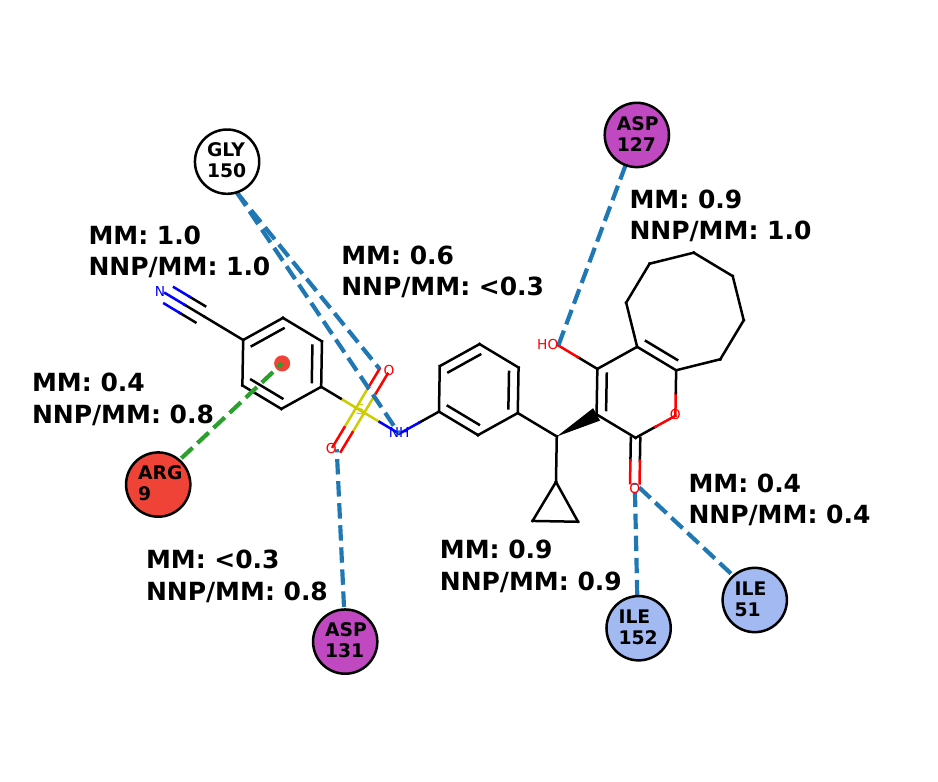}
        \vspace{-0.5cm}
        \caption{1HPO}
    \end{subfigure}\\
    \begin{subfigure}{0.4\textwidth}
        \vspace{-0.5cm}
        \includegraphics[width=\textwidth]{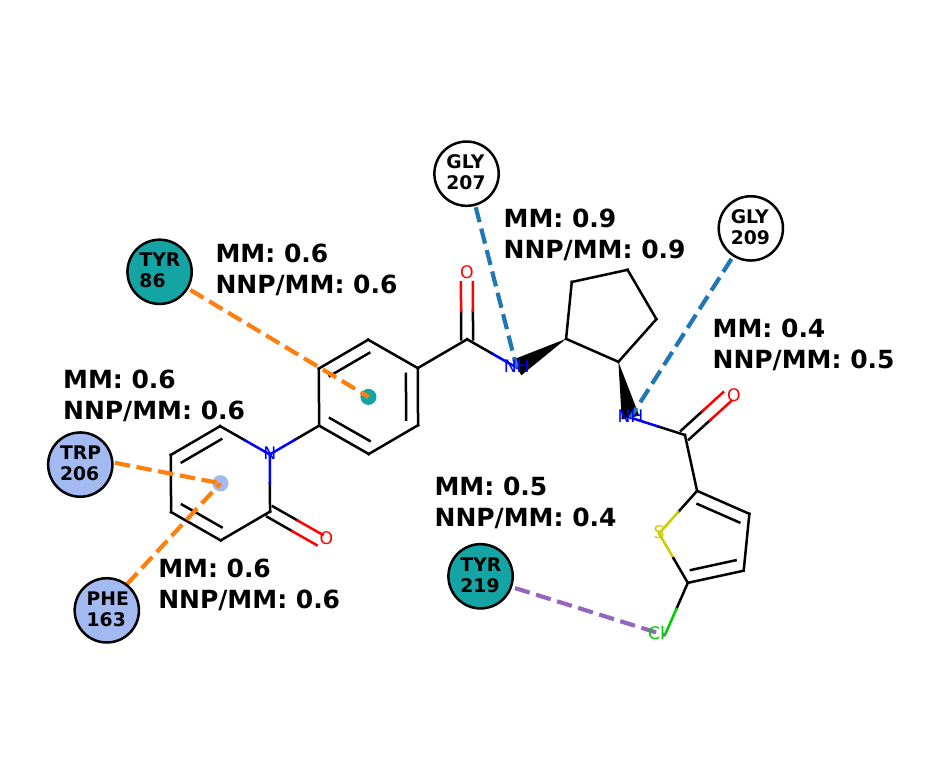}
        \vspace{-1cm}
        \caption{2P95}
    \end{subfigure}
    \begin{subfigure}{0.4\textwidth}
        \vspace{-0.5cm}
        \includegraphics[width=\textwidth]{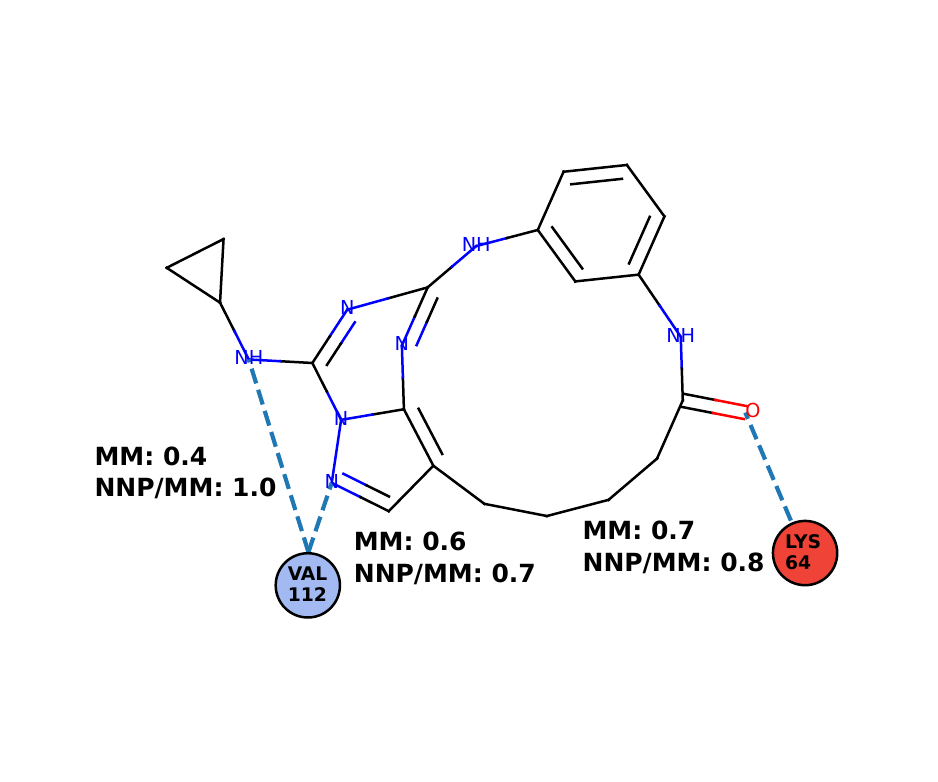}
        \vspace{-1cm}
        \caption{3BE9}
    \end{subfigure}
    \caption{Probabilities of protein-ligand interactions for the different systems and methods (a-d). Protein residues are represented as disks (blue: non-polar, green: polar, red:positively charged, magenta: negatively charged, and dark cyan: aromatic). Interactions are depicted as dashed lines (blue: hydrogen bond, green: {cation-$\pi$}, orange: {$\pi$-$\pi$} interaction, and violet: {$\sigma$-hole}). The interactions with probabilities lower than 0.3 are excluded.}
    \label{fig:plewview}
\end{figure*}

%\todo[inline]{Protein-ligand interactions}
The dominant protein-ligand interactions (Figure~\ref{fig:plewview}) qualitatively agree between MM and NNP/MM for all the complexes.
The full list of ligand-protein interactions and technical details are available in the supplementary information (Table~S1-S8).
Note, the protein-ligand systems have been chosen for demonstration only and further analysis is beyond the scope of this work.

\subsubsection{Simulation speed}

%\todo[inline]{Speed up of NNP}
On average, NNPOps\cite{nnpops} accelerates ANI calculations (energy and forces) 6.5~times (Table~\ref{tab:nnp-speed}).
The is no strict dependency between the ligand size and the calculation time, which suggests significant overhead is coming from auxiliary operations rather than the computation of NNPs.
The overhead mainly comes from PyTorch, which is optimized for batch computing rather than low latency\cite{paszke2019pytorch}.

\begin{table*}[tb]
\caption{Comparison of ANI-2x inference (energy and forces) time (ms) using the original TorchANI and the TorchANI accelerated with NNPOps (TorchANI/NNPOps).
The results were obtained with an NVIDIA RTX 4090 GPU.}
\label{tab:nnp-speed}
\begin{threeparttable}
\begin{tabular}{cccc}
\toprule
System & TorchANI & TorchANI/NNPOps & Speed-up \\
\midrule
1AJV &  11.5 &  2.17 &   5.3 \\
1HPO &  11.3 &  1.91 &   5.9 \\
2P95 &  13.0 &  1.54 &   8.4 \\
3BE9 &   9.4 &  1.52 &   6.2 \\
\midrule
Average &    &       &   6.5 \\
\bottomrule
\end{tabular}
\end{threeparttable}
\end{table*}

%\todo[inline]{Speed up of NNP/MM}
Overall NNP/MM is sped up 5.3~times (Table~\ref{tab:md-speed}) on average when NNPOps\cite{nnpops} is used.
Despite this improvement, NNP/MM is still about an order of magnitude slower than the conventional MM (Table~\ref{tab:md-speed}), but further optimizations are possible.
First, ANI-2x\cite{devereux2020extending} uses an ensemble of 8~NNs.
If only one NN could be used, the simulations would be 2.2~times faster on average (Table~\ref{tab:md-speed}).
Second, the time step for the NNP/MM simulations has to be reduced from \SI{4}{fs} to \SI{2}{fs}.
If the constraint scheme could be adapted to allow \SI{4}{fs} timestep, the simulations would be 2~times faster.
Finally, not all the software components are already fully optimized.
For example, the current implementation of OpenMM-Torch (\url{https://github.com/openmm/openmm-torch}) performs the NNP and MM calculations on a GPU sequentially, but it would be more efficient to do that concurrently.

\begin{table*}[tb]
\caption{Comparison of MD simulation speed (ns/day) of NNP/MM using the original TorchANI, the TorchANI accelerated with NNPOps (TorchANI/NNPOps), and the TorchANI accelerated with NNPOps and just one model of ANI-2x (1~NN).
For reference, MM speed is included.
The results were obtained with an NVIDIA RTX 4090 GPU.}
\label{tab:md-speed}
\begin{threeparttable}
\begin{tabular}{c
                >{\centering}p{0.15\textwidth}
                >{\centering}p{0.25\textwidth}
                >{\centering}p{0.15\textwidth}
                c}
\toprule
System & NNP/MM (TorchANI)\tnote{*} & NNP/MM (TorchANI/NNPOps)\tnote{*} & NNP/MM (1~NN)\tnote{*} & MM\tnote{$\dagger$} \\
\midrule
1AJV & 12.6 & 60.1 & 155 & 1382 \\
% NNPOps speed uP: 4.77
% NN1 speed up: 2.58
1HPO & 13.4 & 65.9 & 152 & 1227 \\
% NNPOps speed uP: 4.92
% NN1 speed up: 2.30
2P95 & 12.2 & 73.5 & 147 & 1006 \\
% NNPOps speed uP:6.02
% NN1 speed up: 2.00
3BE9 & 14.0 & 74.2 & 151 &  995 \\
% NNPOps speed uP: 5.30
% NN1 speed up: 2.04

% Average NNPOps speed uP: 5.25
% Average NN1 speed up: 2.23
\bottomrule
\end{tabular}
\begin{tablenotes}
\item[*] \SI{2}{fs} time step
\item[$\dagger$] \SI{4}{fs} time step
\end{tablenotes}
\end{threeparttable}
\end{table*}

\subsection{Extensibility with other NNPs}

%\todo[inline]{ACEMD can work with other NNPs}
Our implementation of NNP/MM is agnostic to the NNP model, i.e. it can use any model implemented with PyTorch.
As a demonstration, we performed simulations with ANI-1x\cite{smith2018less} and TorchMD-NET\cite{tholke2022equivariant} trained with the ANI-1 data set\cite{smith2017anidata}.
The simulation speed benchmarks (Table~\ref{tab:md-speed-various}) are available just for 3BE9 because both NNPs are limited to 4 elements (H, C, N, and O).

\begin{table}
\caption{Comparison of MD simulation speed of ANI-1x and TorchMD-NET and their accuracy in mean absolute error (MAE). The results were obtained with an NVIDIA RTX 4090 GPU. }
\label{tab:md-speed-various}
\begin{threeparttable}
\begin{tabular}{c
                >{\centering}p{0.08\textwidth}
                >{\centering}p{0.1\textwidth}
                c}
\toprule
System 3BE9 & ANI-1x\tnote{*}  & TorchMD-NET\tnote{*} & \\
\midrule
speed (ns/day) & 127 & 17.0 & \\
accuracy (eV) & 0.057 & 0.010 & \\
\bottomrule
\end{tabular}
\begin{tablenotes}
\item[*] \SI{2}{fs} time step
\end{tablenotes}
\end{threeparttable}
\end{table}

\section{Software installation and usage}

%\todo[inline]{Installation}
ACEMD can be installed with the Conda package management system\cite{conda}.
For dependencies, Conda-forge\cite{conda-forge} is used to ensure compatibility with all major Linux distributions (refer to the ACEMD documentation for details at \url{https://software.acellera.com}.
The installation command:
\begin{verbatim}
$ conda install -c conda-forge \
                -c acellera \
                -c acellera/label/rc \
                acemd=4
\end{verbatim}
For the best performance, it is recommended to have an NVIDIA GPU and its latest drivers installed, but it is possible to run on a CPU only. 

%\todo[inline]{Setup}
The setup of an NNP/MM simulation consists of the following steps.
First, a system needs to be prepared for a conventional MM simulation  (i.e. initial structure, topology, and force field parameters).
Note that the NNP atoms need to be assigned partial charges and Lennard-Jones parameters to compute the coupling term correctly.
The system preparation can be easily accomplished with HTMD\cite{doerr2016htmd,htmd}.
Second, NNP model files need to be generated with \verb"prepare-nnp" tool included with ACEMD.
It needs the initial structure (e.g. \verb"structure.pdb"), a selection of the NNP atoms (e.g. \verb'"resname MOL"'), and a name of NNP 
\begin{verbatim}
$ prepare-nnp structure.pdb --selection "resname MOL" \
                            --model ANI-2x
\end{verbatim}
The tool generates several files including \verb"model.json".
Currently, we plan to support the NNP models from TorchANI\cite{gao2020torchani} and TorchMD-NET\cite{tholke2022equivariant} but other models will also be supported in the future.
Finally, an ACEMD input file needs to be prepared as for a conventional MD simulation (refer to the ACEMD documentation\cite{acemd} for details) and needs just one additional line (\verb"nnpfile model.json") to enable NNP/MM.

\section{Conclusion}
We have showcased an optimized implementation of NNP/MM in ACEMD\cite{harvey2009acemd}, based on OpenMM\cite{eastman2010openmm} and PyTorch\cite{paszke2019pytorch}, which delivers simulation speeds of approximately 5 times faster than previously reported.
While still slower than classical force fields, the enhanced accuracy of NNPs may justify the increased computational expense (see \citeauthor{rufa2020towards}\cite{rufa2020towards}).
We anticipate this performance gap will continue to shrink in the future.
Presently, NNPs have limited applicability due to constraints on charges and elements, but improvements are expected in the near future.

We validated our implementation by conducting metadynamics simulations of an erlotinib fragment and molecular dynamics simulations of four protein-ligand complexes.
The fragment simulation results are consistent with prior findings, while the complex simulations exceeded previous durations by over an order of magnitude.
These outcomes confirm the effectiveness of our implementation and demonstrate its practical application.
Furthermore, NNP/MM can be combined with the enhanced sampling methods (e.g. metadynamics\cite{barducci2011metadynamics}, replica exchange\cite{sugita1999replica}, steered molecular dynamics\cite{izrailev1999steered}, etc.) and it holds significant potential for alchemical free energy simulations\cite{rufa2020towards}.
It is particularly beneficial for drug discovery efforts, where the simulation of novel molecules is routine but accurate force field parameters may be lacking.

\begin{acknowledgement}
The authors thank the volunteers of GPUGRID.net for donating computing time.
This project has received funding from
the Torres-Quevedo Programme from the Spanish National Agency for Research (PTQ-17-09078 / AEI / 10.13039/501100011033) (RG);
the European Union’s Horizon 2020 research and innovation programme under grant agreement No. 823712 (RG, AV, RF);
the Industrial Doctorates Plan of the Secretariat of Universities and Research of the Department of Economy and Knowledge of the Generalitat of Catalonia (AV);
the Chan Zuckerberg Initiative DAF (grant number: 2020-219414), an advised fund of Silicon Valley Community Foundation (SD, PE);
and the project PID2020-116564GB-I00 has been funded by MCIN / AEI / 10.13039/501100011033.
Research reported in this publication was supported by the National Institute of General Medical Sciences (NIGMS) of the National Institutes of Health under award number GM140090 (PE, TEM, JDC, GDF).
The content is solely the responsibility of the authors and does not necessarily represent the official views of the National Institutes of Health.
JDC is a current member of the Scientific Advisory Board of OpenEye Scientific Software, Redesign Science, Ventus Therapeutics, and Interline Therapeutics, and has equity interests in Redesign Science and Interline Therapeutics. The Chodera laboratory receives or has received funding from multiple sources, including the National Institutes of Health, the National Science Foundation, the Parker Institute for Cancer Immunotherapy, Relay Therapeutics, Entasis Therapeutics, Silicon Therapeutics, EMD Serono (Merck KGaA), AstraZeneca, Vir Biotechnology, Bayer, XtalPi, Interline Therapeutics, the Molecular Sciences Software Institute, the Starr Cancer Consortium, the Open Force Field Consortium, Cycle for Survival, a Louis V. Gerstner Young Investigator Award, and the Sloan Kettering Institute.
A complete funding history for the Chodera lab can be found at \url{http://choderalab.org/funding}.
\end{acknowledgement}

\begin{suppinfo}
The following files are available free of charge:
\begin{itemize}
  \item SI.pdf: the time series of the dihedral angles of the fragment; the energy profiles of the dihedral angle scan of the ligands; the protein and ligand RMSD and residue RMSF for all the simulations; and the full list of ligand-protein interactions.
\end{itemize}
Installation instructions for the software are available at \url{https://software.acellera.com}.
\end{suppinfo}

\bibliography{references}

\begin{mcitethebibliography}{72}
\bibitem[Lin and Truhlar(2007)Lin, and Truhlar]{lin2007qm}
Lin,~H.; Truhlar,~D.~G. QM/MM: what have we learned, where are we, and where do
  we go from here? \emph{Theor. Chem. Acc} \textbf{2007}, \emph{117},
  185--199\relax
\mciteBstWouldAddEndPuncttrue
\mciteSetBstMidEndSepPunct{\mcitedefaultmidpunct}
{\mcitedefaultendpunct}{\mcitedefaultseppunct}\relax
\EndOfBibitem
\end{mcitethebibliography}

\end{document}

\clearpage
\section{Preparation and simulation conditions for the fragment system}

%\todo[inline]{System preparation}
The system preparation has been carried out with HTMD\cite{doerr2016htmd}.
The fragment is constructed from SMILES and solvated with \SI{5}{\angstrom} padding.

%\todo[inline]{FF and MD conditions}
The molecular topology and FF parameters files have been prepared with HTMD\cite{doerr2016htmd}.
The force field parameters are obtained with GAFF2\cite{wang2004development} or modeled with ANI-2x\cite{devereux2020extending} depending on the simulation method.
For water, the rigid TIP3P\cite{jorgensen1983comparison} model is used.
The bonds between hydrogens and heavy atoms (\ch{H-X}) are constrained and the hydrogen masses are repartitioned to \SI{4.0}{u}.
The electrostatic interactions are computed with PME\cite{essmann1995smooth} and the van der Waals interactions with the cutoff of \SI{9.0}{\angstrom}.
The MD time step for the MM simulations is \SI{4.0}{fs}, but for NNP/MM it is reduced to \SI{2.0}{fs} due to stability issues.
The temperature and pressures are maintained with the Langevin thermostat (the damping constant of \SI{0.1}{ps^{-1}}) and the isotropic Monte Carlo barostat, respectively.

%\todo[inline]{MTD setup}
For metadynamics (MTD), PLUMED\cite{tribello2014plumed} plugin is used.
Two dihedral angles are used as collective variables.
The bias potential is updated every \SI{1}{ps}, kernel height is \SI{4.185}{kcal/mol} and width \SI{10}{\deg}.
The bias factor is set to 5 (i.e. the effective temperature is \SI{1550}{K})\cite{barducci2008well}.

%\todo[inline]{Equilibration and production}
Each protein-ligand complex has been equilibrated with the following procedure.
First, the system is minimized for 500~steps.
Second, the system is equilibrated for \SI{1.0}{ns} with the NPT ensemble (T~=~\SI{310}{K} and P~=~\SI{1.0}{bar}).
For the production simulations, the NVT ensemble (T~=~\SI{310}{K}) is used without any restraints.
Simulation trajectories are saved every \SI{50}{ps}.

\clearpage
\section{Dihedral angle series of the fragment}

\begin{figure}
    \centering
    \includegraphics[width=\textwidth]{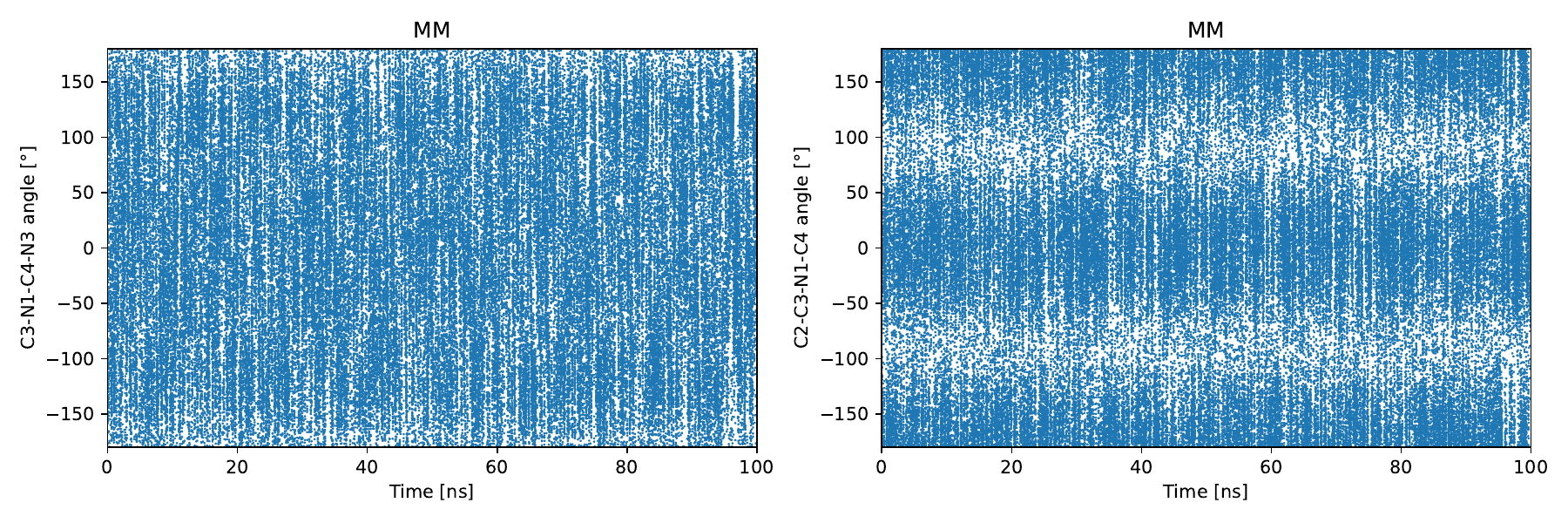}
    \caption{The dihedral angle series of the fragment from the MM simulation.}
\end{figure}
\begin{figure}
    \centering
    \includegraphics[width=\textwidth]{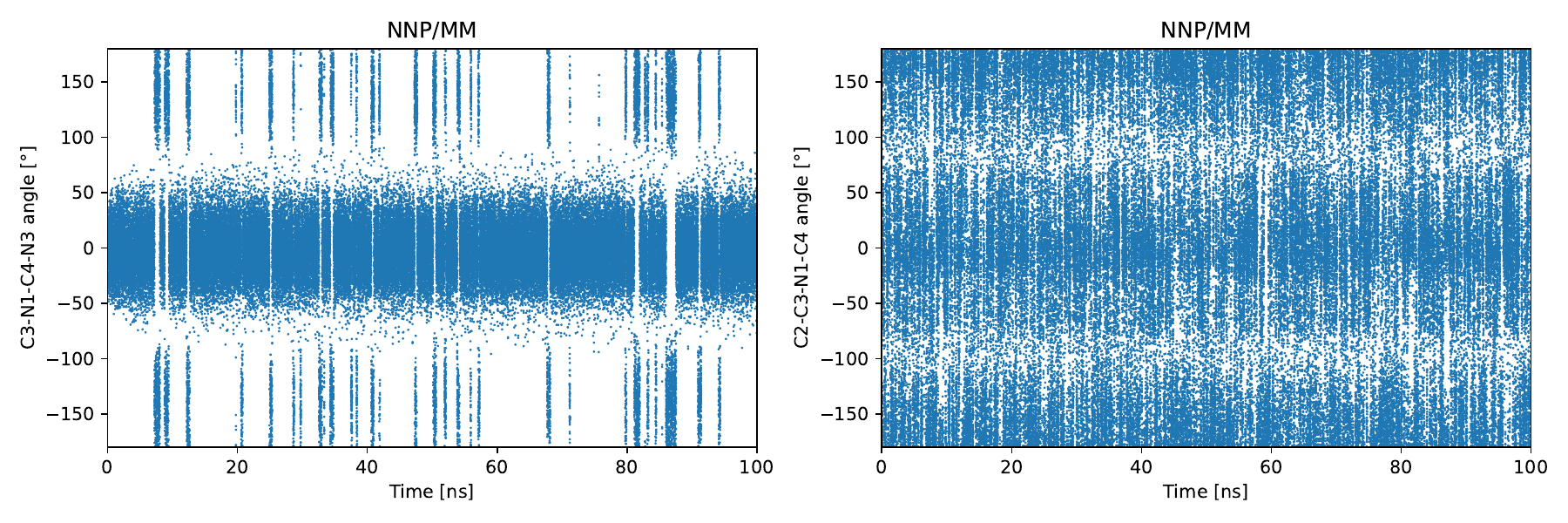}
    \caption{The dihedral angle series of the fragment from the NNP/MM simulation.}
\end{figure}

\clearpage
\section{Preparation and simulation conditions for protein-ligand systems}

%\todo[inline]{System preparation}
The system preparation has been carried out with HTMD\cite{doerr2016htmd}.
First, the structure of a protein-ligand complex is downloaded from the PDB\cite{berman2003announcing,wwpdb2019protein} database.
Only the protein and ligand are kept, and the other atoms are removed.
Second, the protein is protonated at pH~=~7.0 with PROPKA\cite{sondergaard2011improved,olsson2011propka3} and its hydrogen bonds are optimized with PDB2PQR\cite{dolinsky2004pdb2pqr,dolinsky2007pdb2pqr}.
Finally, the system is solvated with \SI{5}{\angstrom} padding and neutralized with \ch{Na+} or \ch{Cl-} ions.

%\todo[inline]{FF and MD conditions}
The molecular topology and FF parameters files have been prepared with HTMD\cite{doerr2016htmd}.
For the protein, AMBER ff14SB\cite{maier2015ff14sb} FF is used.
For the ligand, the force field parameters are obtained with GAFF2\cite{wang2004development} or modeled with ANI-2x\cite{devereux2020extending} depending on the simulation method.
For water, the rigid TIP3P\cite{jorgensen1983comparison} model is used.
The bonds between hydrogens and heavy atoms (\ch{H-X}) are constrained and the hydrogen masses are repartitioned to \SI{4.0}{u}.
The electrostatic interactions are computed with PME\cite{essmann1995smooth} and the van der Waals interactions with the cutoff of \SI{9.0}{\angstrom}.
The MD time step for the MM simulations is \SI{4.0}{fs}, but for NNP/MM it is reduced to \SI{2.0}{fs} due to stability issues.
The temperature and pressures are maintained with the Langevin thermostat (the damping constant of \SI{0.1}{ps^{-1}}) and the isotropic Monte Carlo barostat, respectively.

%\todo[inline]{Equilibration and production}
Each protein-ligand complex has been equilibrated with the following procedure.
First, the system is minimized for 500~steps with the harmonic positional restraint applied to the protein (the force constant for backbone atoms is \SI{1.0}{kcal/mol/\angstrom^2} and the one for the side-chain atoms is \SI{1.0}{kcal/mol/\angstrom^2}).
Second, the system is equilibrated for \SI{1.0}{ns} with the NPT ensemble (T~=~\SI{310}{K} and P~=~\SI{1.0}{bar}).
The force constants of the restraints are reduced linearly to \SI{1.0}{kcal/mol/\angstrom^2} during the first \SI{0.5}{ns} of the simulation.
For the production simulations, the NVT ensemble (T~=~\SI{310}{K}) is used without any restraints.
Simulation trajectories are saved every \SI{50}{ps}.

\clearpage
\section{Dihedral angle scans}

\begin{figure}
    \centering
    \includegraphics[width=0.33\textwidth]{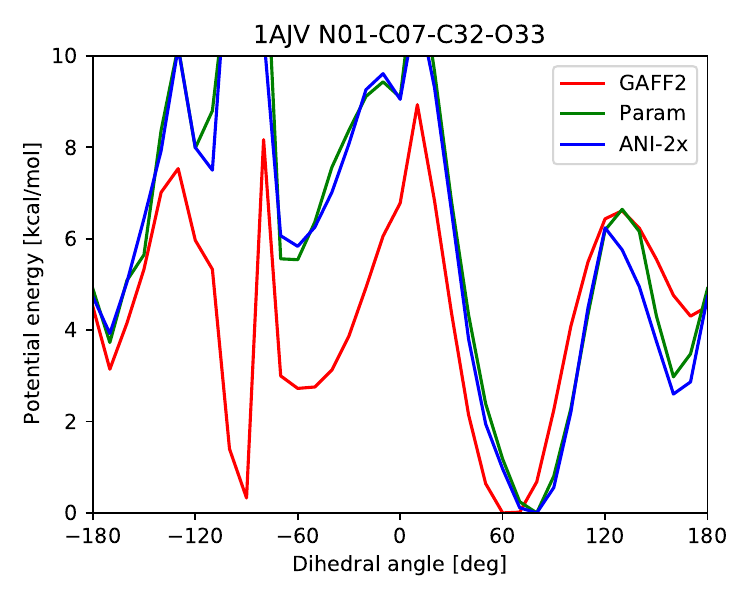}
    \caption{Energy profile of the dihedral angle N01-C07-C32-O33 scan of 1AJV ligand. Rotamer energies are computed with different methods.}
\end{figure}
\begin{figure}
    \centering
    \includegraphics[width=0.33\textwidth]{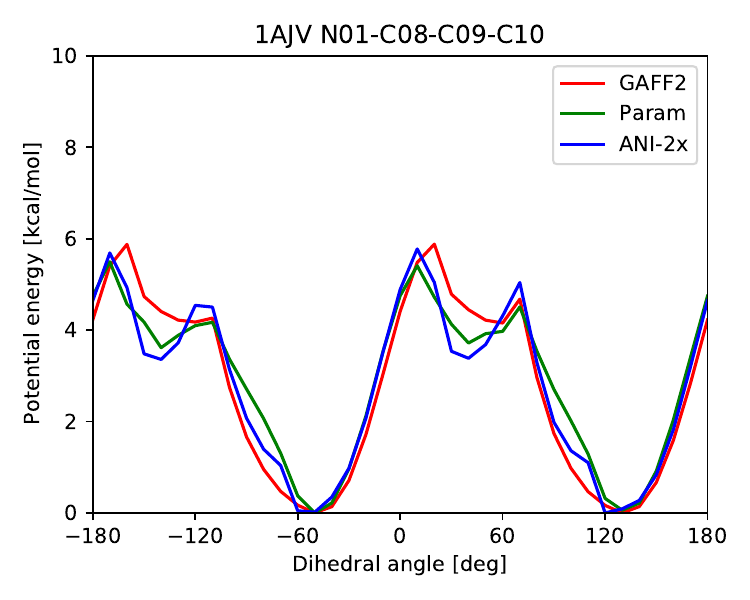}
    \caption{Energy profile of the dihedral angle N01-C08-C09-C10 scan of 1AJV ligand. Rotamer energies are computed with different methods.}
\end{figure}
\begin{figure}
    \centering
    \includegraphics[width=0.33\textwidth]{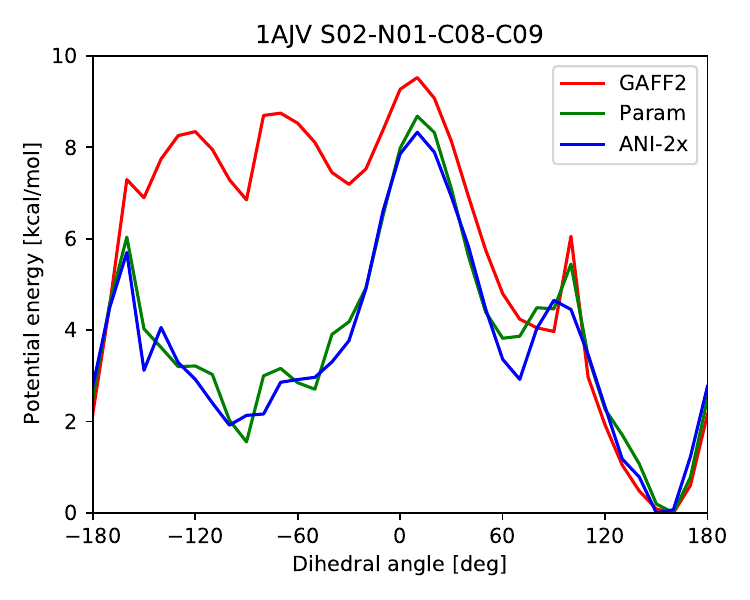}
    \caption{Energy profile of the dihedral angle S02-N01-C08-C09 scan of 1AJV ligand. Rotamer energies are computed with different methods.}
\end{figure}
\begin{figure}
    \centering
    \includegraphics[width=0.33\textwidth]{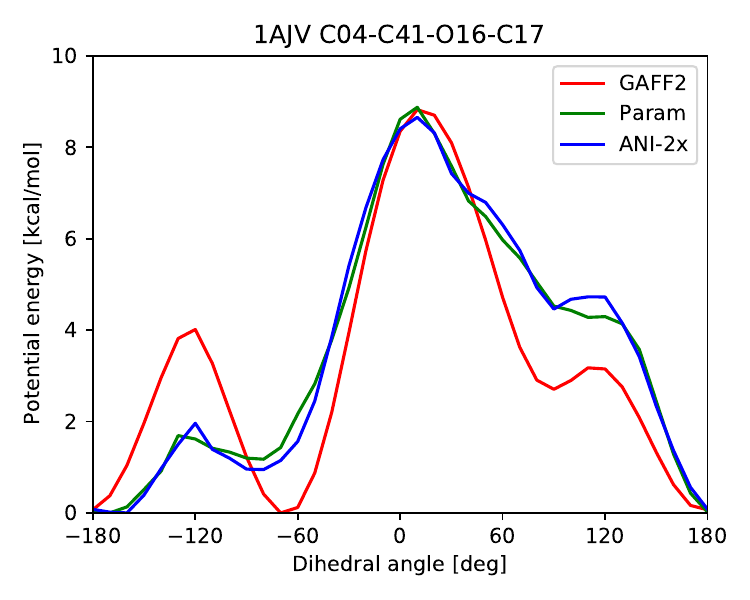}
    \caption{Energy profile of the dihedral angle C04-C41-O16-C17 scan of 1AJV ligand. Rotamer energies are computed with different methods.}
\end{figure}
\begin{figure}
    \centering
    \includegraphics[width=0.33\textwidth]{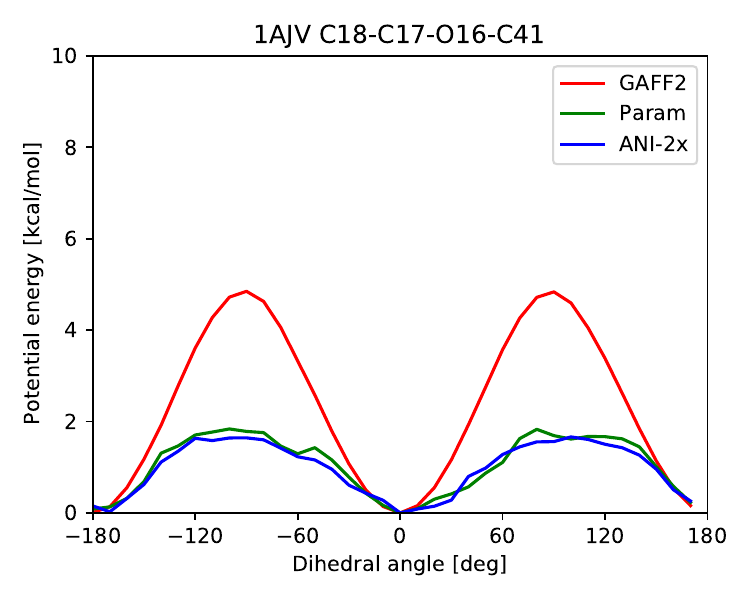}
    \caption{Energy profile of the dihedral angle C18-C17-O16-C41 scan of 1AJV ligand. Rotamer energies are computed with different methods.}
\end{figure}
\begin{figure}
    \centering
    \includegraphics[width=0.33\textwidth]{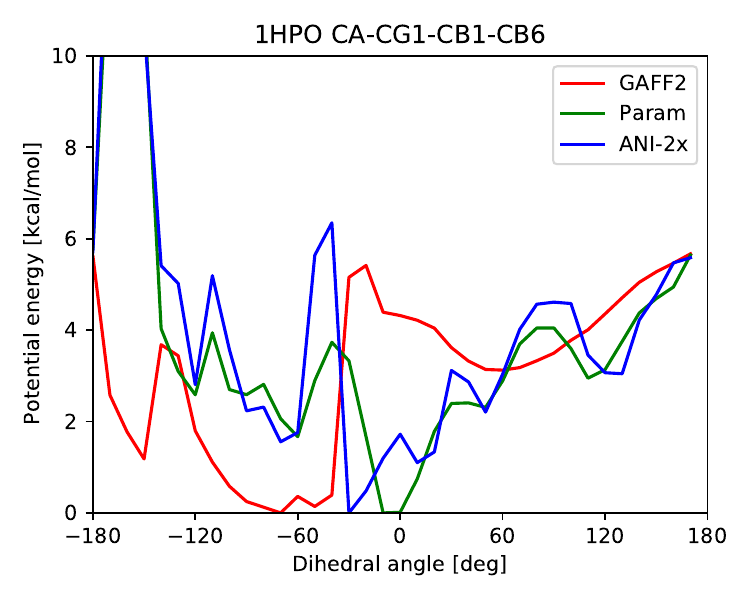}
    \caption{Energy profile of the dihedral angle CA-CG1-CB1-CB6 scan of 1HPO ligand. Rotamer energies are computed with different methods.}
\end{figure}
\begin{figure}
    \centering
    \includegraphics[width=0.33\textwidth]{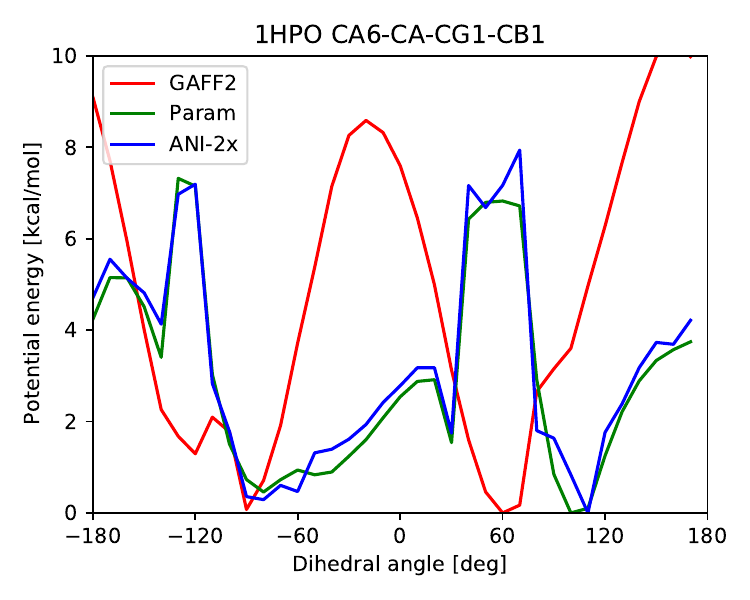}
    \caption{Energy profile of the dihedral angle CA6-CA-CG1-CB1 scan of 1HPO ligand. Rotamer energies are computed with different methods.}
\end{figure}
\begin{figure}
    \centering
    \includegraphics[width=0.33\textwidth]{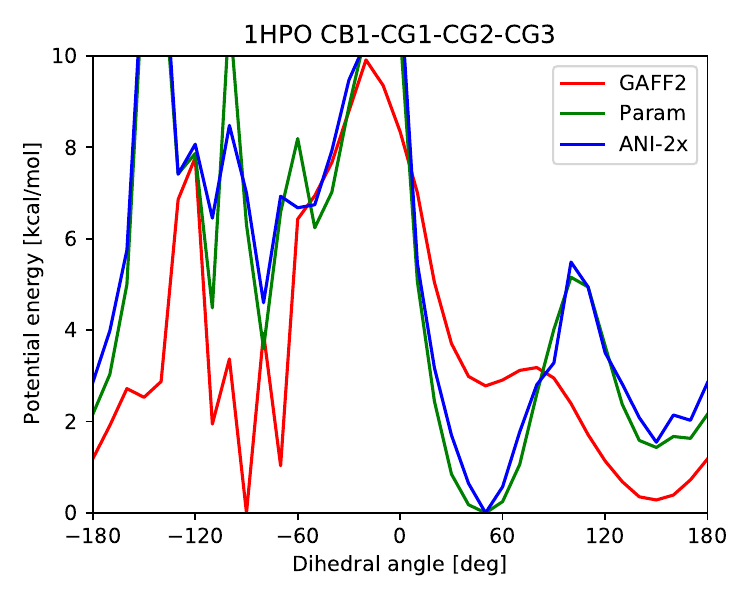}
    \caption{Energy profile of the dihedral angle CB1-CG1-CG2-CG3 scan of 1HPO ligand. Rotamer energies are computed with different methods.}
\end{figure}
\begin{figure}
    \centering
    \includegraphics[width=0.33\textwidth]{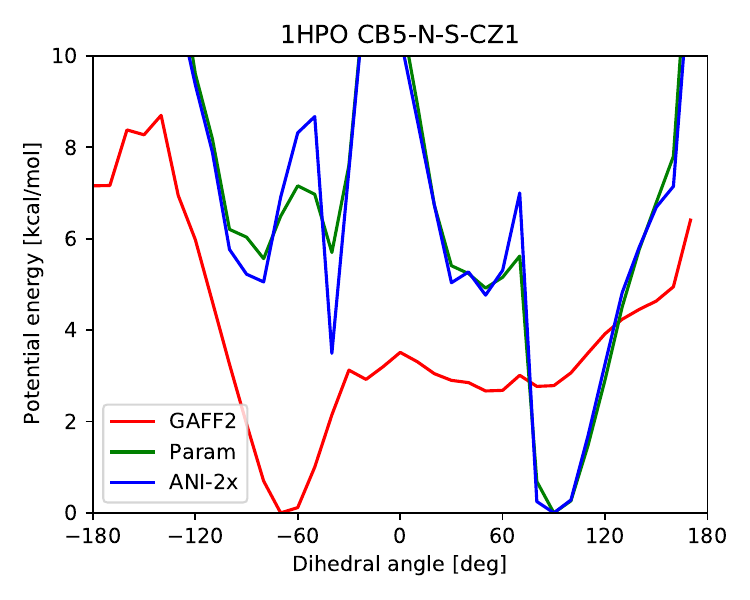}
    \caption{Energy profile of the dihedral angle CB5-N-S-CZ1 scan of 1HPO ligand. Rotamer energies are computed with different methods.}
\end{figure}
\begin{figure}
    \centering
    \includegraphics[width=0.33\textwidth]{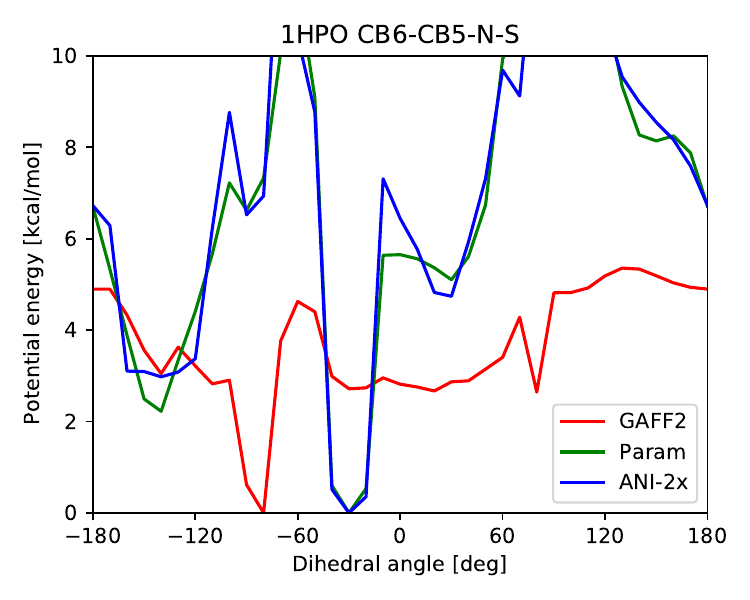}
    \caption{Energy profile of the dihedral angle CB6-CB5-N-S scan of 1HPO ligand. Rotamer energies are computed with different methods.}
\end{figure}
\begin{figure}
    \centering
    \includegraphics[width=0.33\textwidth]{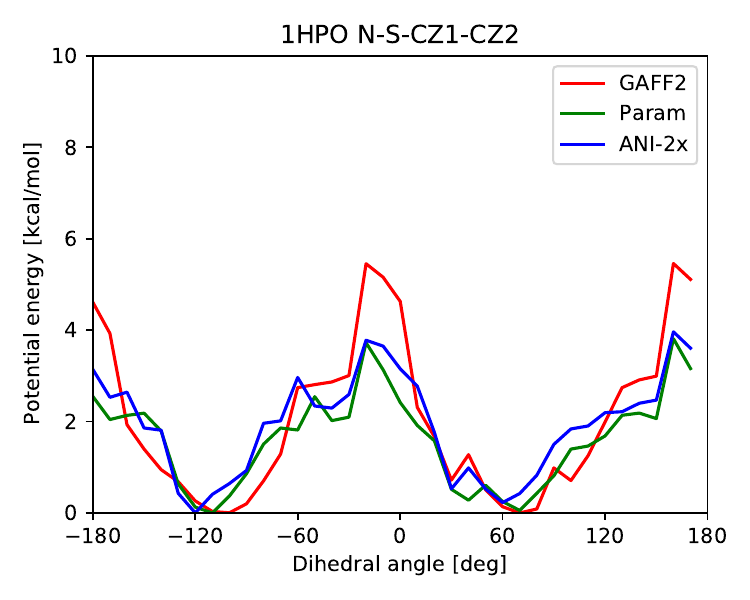}
    \caption{Energy profile of the dihedral angle N-S-CZ1-CZ2 scan of 1HPO ligand. Rotamer energies are computed with different methods.}
\end{figure}
\begin{figure}
    \centering
    \includegraphics[width=0.33\textwidth]{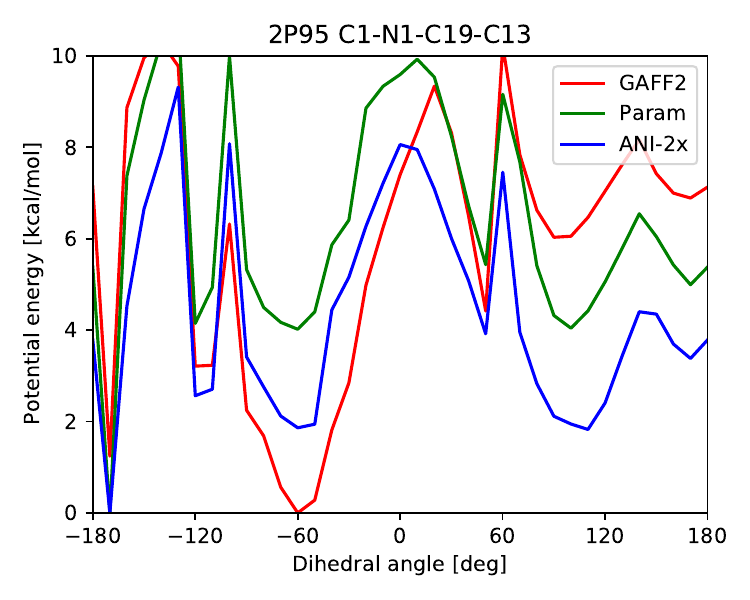}
    \caption{Energy profile of the dihedral angle C1-N1-C19-C13 scan of 2P95 ligand. Rotamer energies are computed with different methods.}
\end{figure}
\begin{figure}
    \centering
    \includegraphics[width=0.33\textwidth]{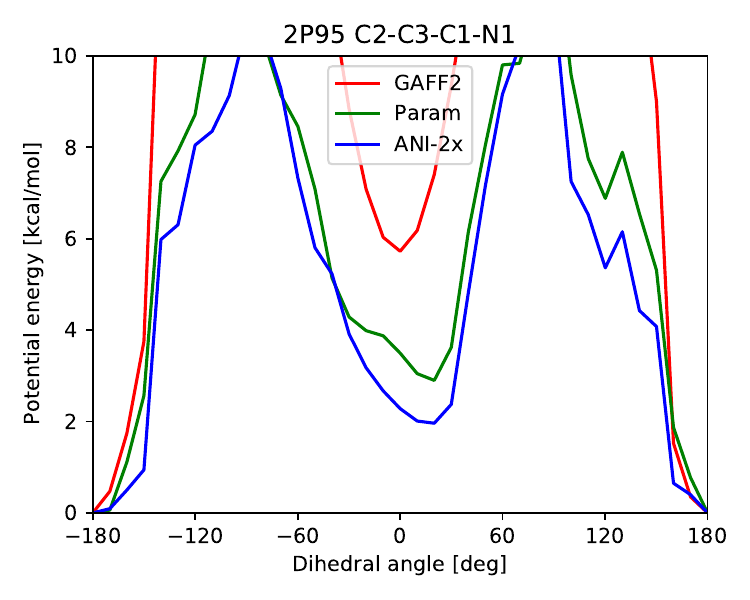}
    \caption{Energy profile of the dihedral angle C2-C3-C1-N1 scan of 2P95 ligand. Rotamer energies are computed with different methods.}
\end{figure}
\begin{figure}
    \centering
    \includegraphics[width=0.33\textwidth]{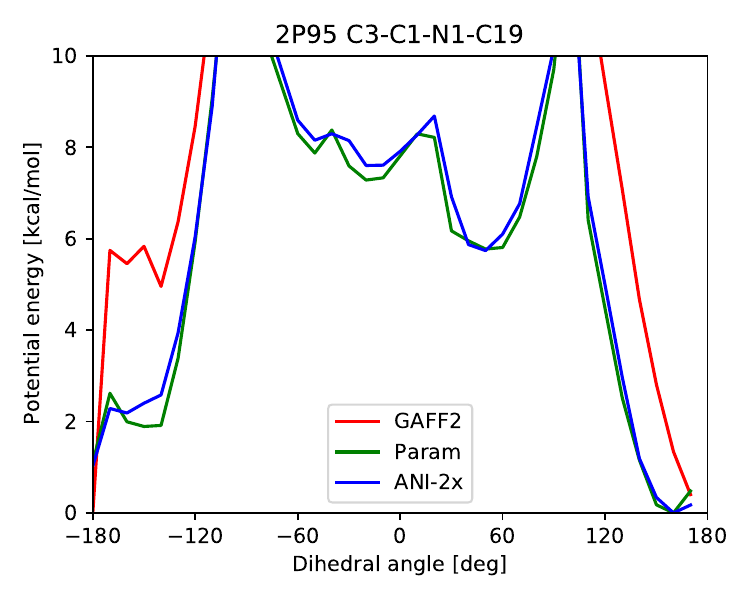}
    \caption{Energy profile of the dihedral angle C3-C1-N1-C19 scan of 2P95 ligand. Rotamer energies are computed with different methods.}
\end{figure}
\begin{figure}
    \centering
    \includegraphics[width=0.33\textwidth]{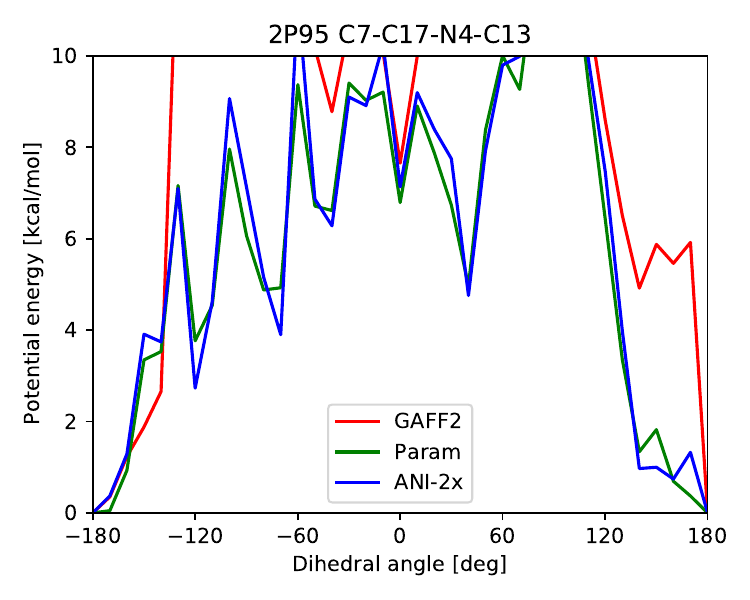}
    \caption{Energy profile of the dihedral angle C7-C17-N4-C13 scan of 2P95 ligand. Rotamer energies are computed with different methods.}
\end{figure}
\begin{figure}
    \centering
    \includegraphics[width=0.33\textwidth]{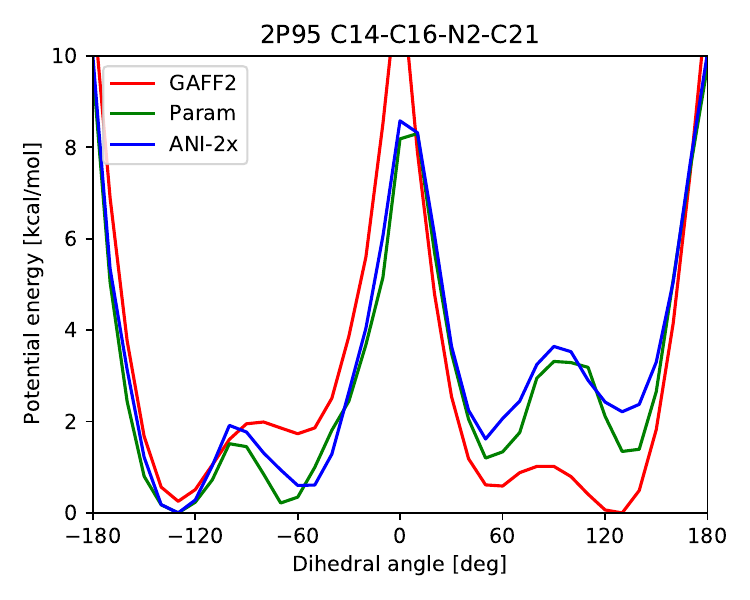}
    \caption{Energy profile of the dihedral angle C14-C16-N2-C21 scan of 2P95 ligand. Rotamer energies are computed with different methods.}
\end{figure}
\begin{figure}
    \centering
    \includegraphics[width=0.33\textwidth]{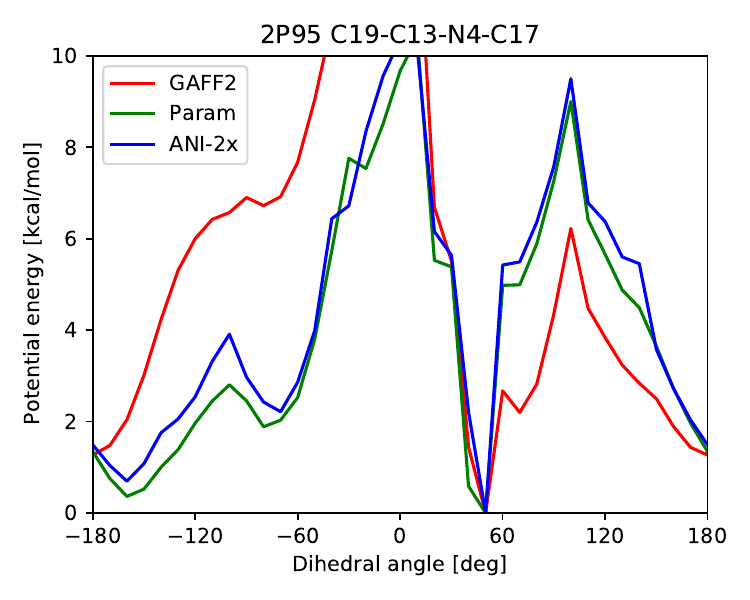}
    \caption{Energy profile of the dihedral angle C19-C13-N4-C17 scan of 2P95 ligand. Rotamer energies are computed with different methods.}
\end{figure}
\begin{figure}
    \centering
    \includegraphics[width=0.33\textwidth]{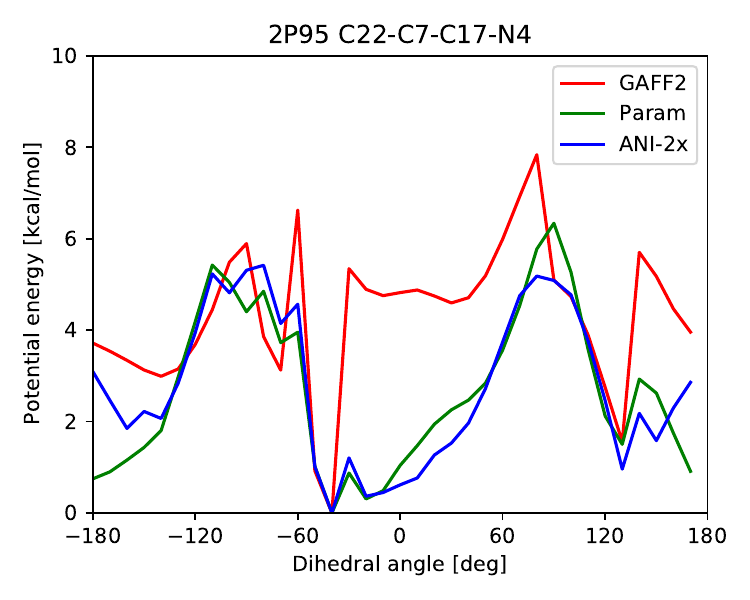}
    \caption{Energy profile of the dihedral angle C22-C7-C17-N4 scan of 2P95 ligand. Rotamer energies are computed with different methods.}
\end{figure}
\begin{figure}
    \centering
    \includegraphics[width=0.33\textwidth]{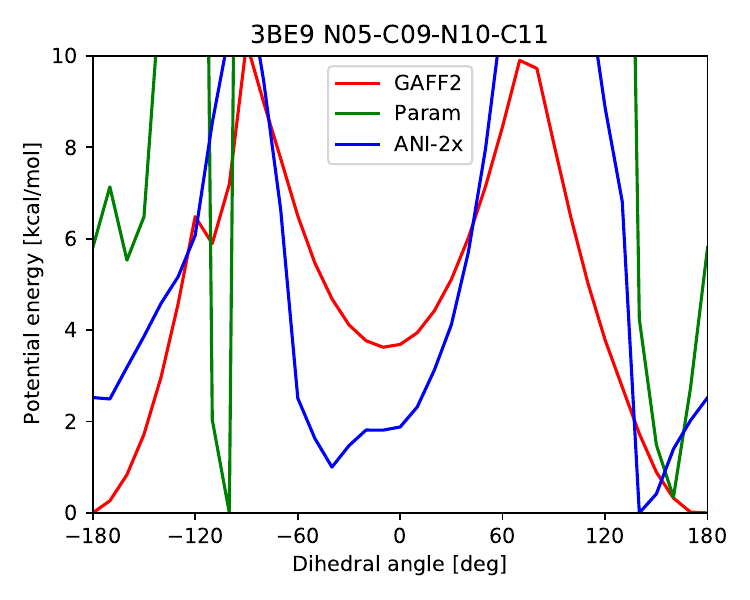}
    \caption{Energy profile of the dihedral angle N05-C09-N10-C11 scan of 3BE9 ligand. Rotamer energies are computed with different methods.}
\end{figure}
\begin{figure}
    \centering
    \includegraphics[width=0.33\textwidth]{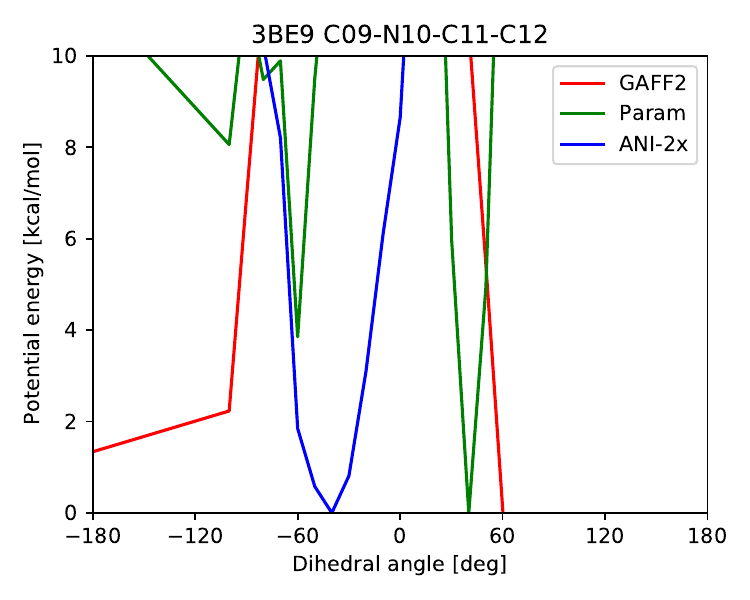}
    \caption{Energy profile of the dihedral angle C09-N10-C11-C12 scan of 3BE9 ligand. Rotamer energies are computed with different methods.}
\end{figure}

\clearpage
\section{Time series of protein RMSD}

\begin{figure}
    \centering
    \hspace{0.05\textwidth}MM\hspace{0.35\textwidth}NNP/MM
    \includegraphics[width=0.9\textwidth]{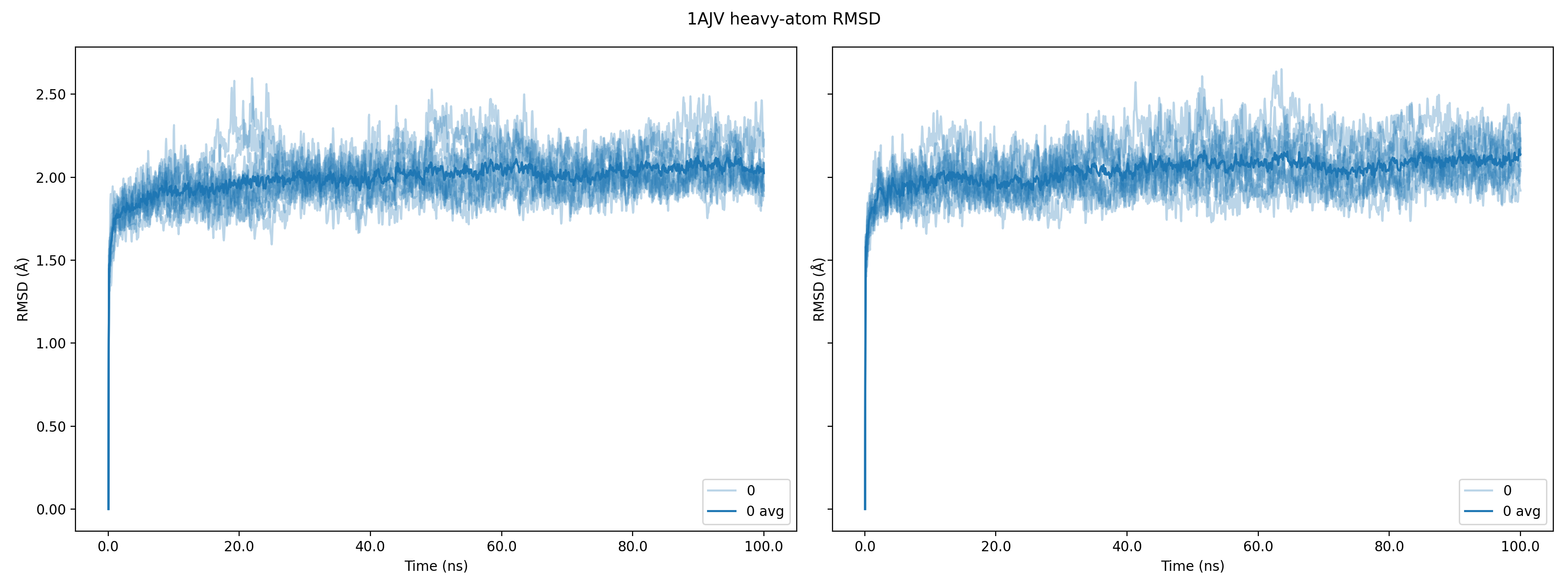}
    \caption{Time series of protein RMSD for 1AJV with the different methods (MM and NNP/MM). The independent simulations are colored in light blue and the average is in dark blue.}
\end{figure}
\begin{figure}
    \centering
    \hspace{0.05\textwidth}MM\hspace{0.35\textwidth}NNP/MM
    \includegraphics[width=0.9\textwidth]{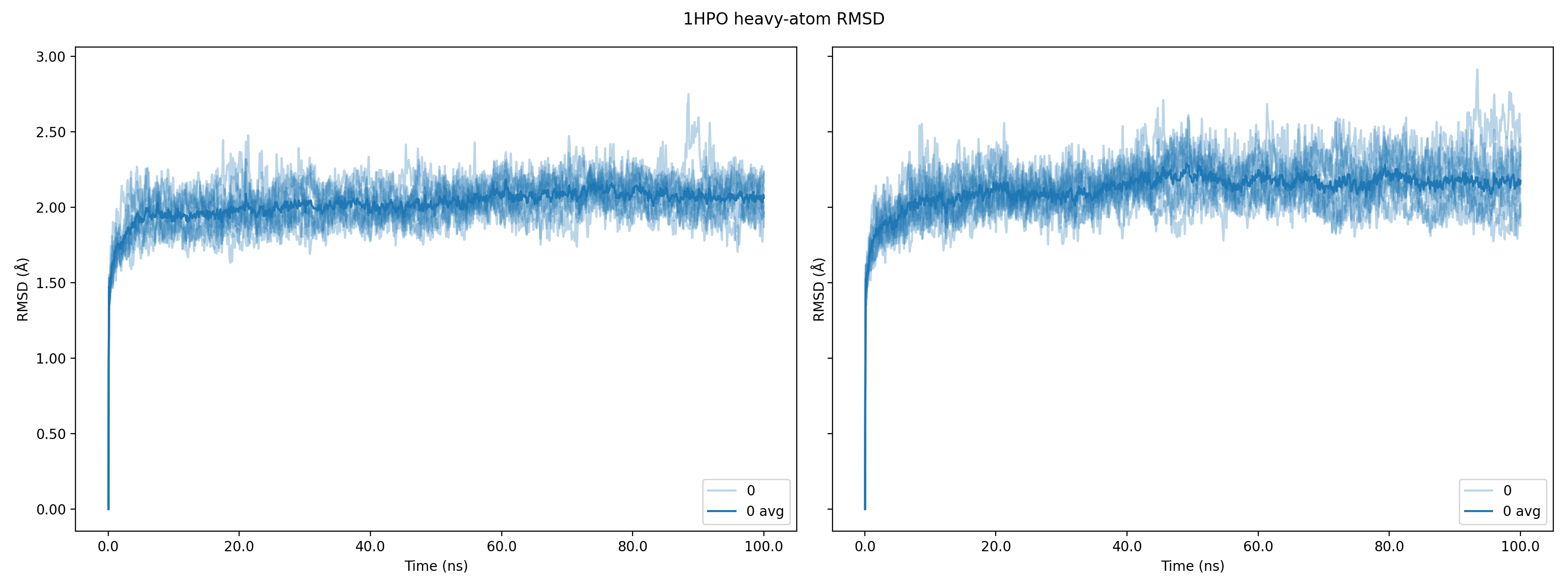}
    \caption{Time series of protein RMSD for 1HPO with the different methods (MM and NNP/MM). The independent simulations are colored in light blue and the average is in dark blue.}
\end{figure}
\begin{figure}
    \centering
    \hspace{0.05\textwidth}MM\hspace{0.35\textwidth}NNP/MM
    \includegraphics[width=0.9\textwidth]{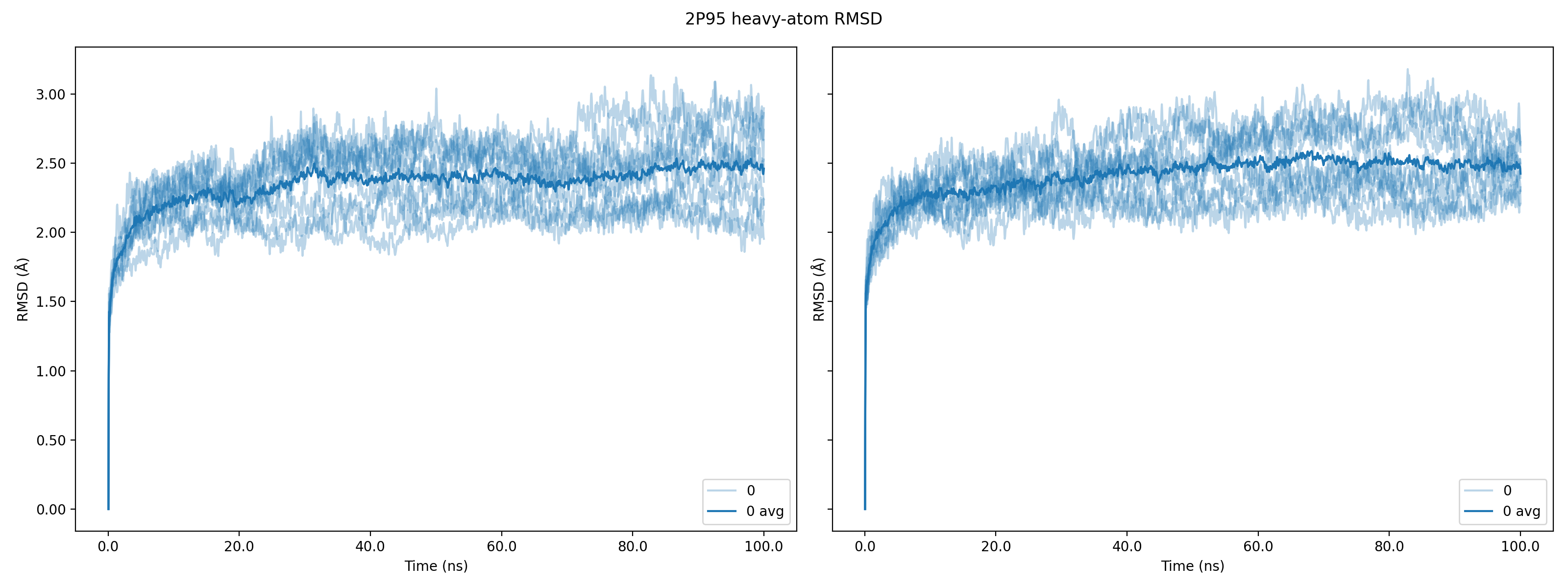}
    \caption{Time series of protein RMSD for 2P95 with the different methods (MM and NNP/MM). The independent simulations are colored in light blue and the average is in dark blue.}
\end{figure}
\begin{figure}
    \centering
    \hspace{0.05\textwidth}MM\hspace{0.35\textwidth}NNP/MM
    \includegraphics[width=0.9\textwidth]{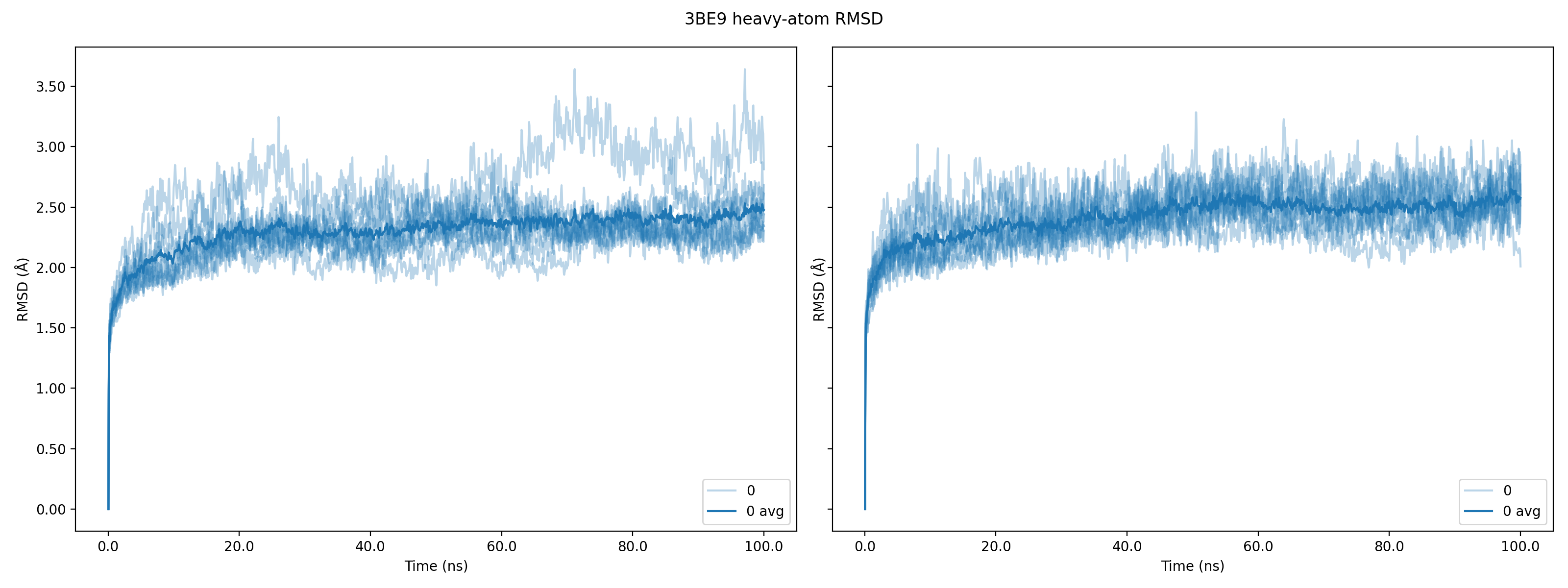}
    \caption{Time series of protein RMSD for 3BE9 with the different methods (MM and NNP/MM). The independent simulations are colored in light blue and the average is in dark blue.}
\end{figure}

\clearpage
\section{Per-residue RMSF of protein}

\begin{figure}
    \centering
    \hspace{0.05\textwidth}MM\hspace{0.35\textwidth}NNP/MM
    \includegraphics[width=0.9\textwidth]{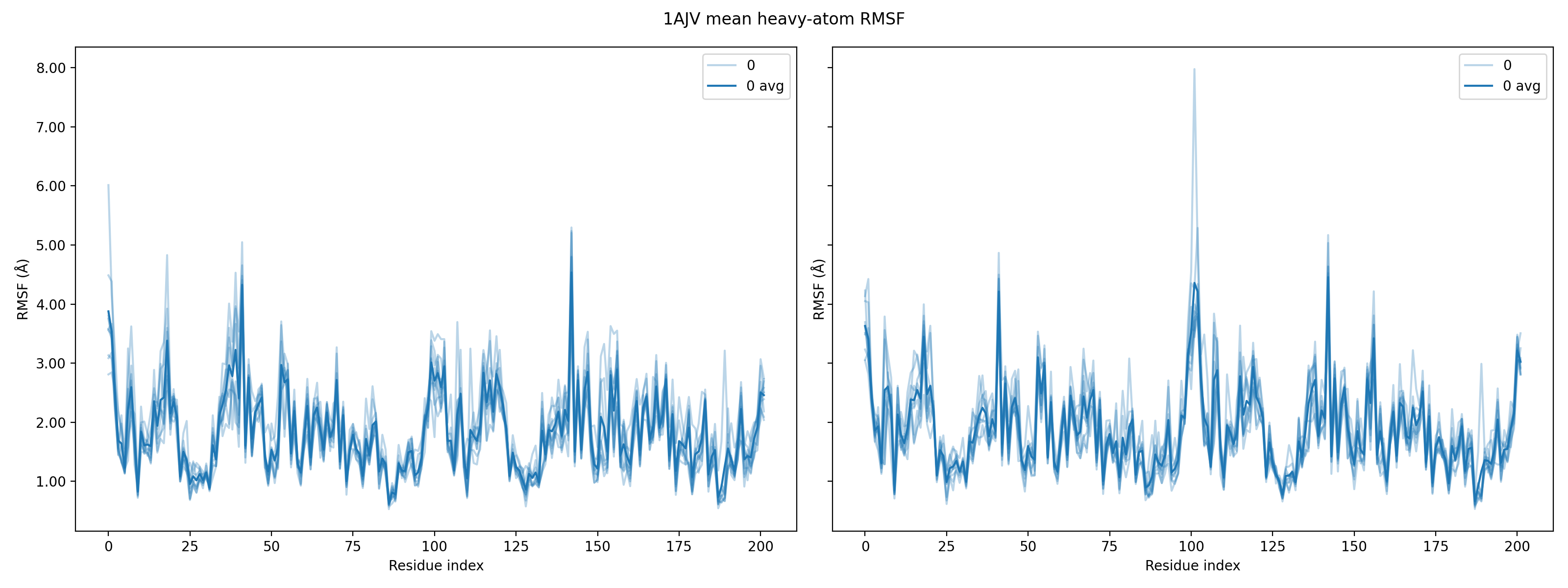}
    \caption{Residue RMSF of protein for 1AJV with the different methods (MM and NNP/MM). The independent simulations are colored in light blue and the average is in dark blue.}
\end{figure}
\begin{figure}
    \centering
    \hspace{0.05\textwidth}MM\hspace{0.35\textwidth}NNP/MM
    \includegraphics[width=0.9\textwidth]{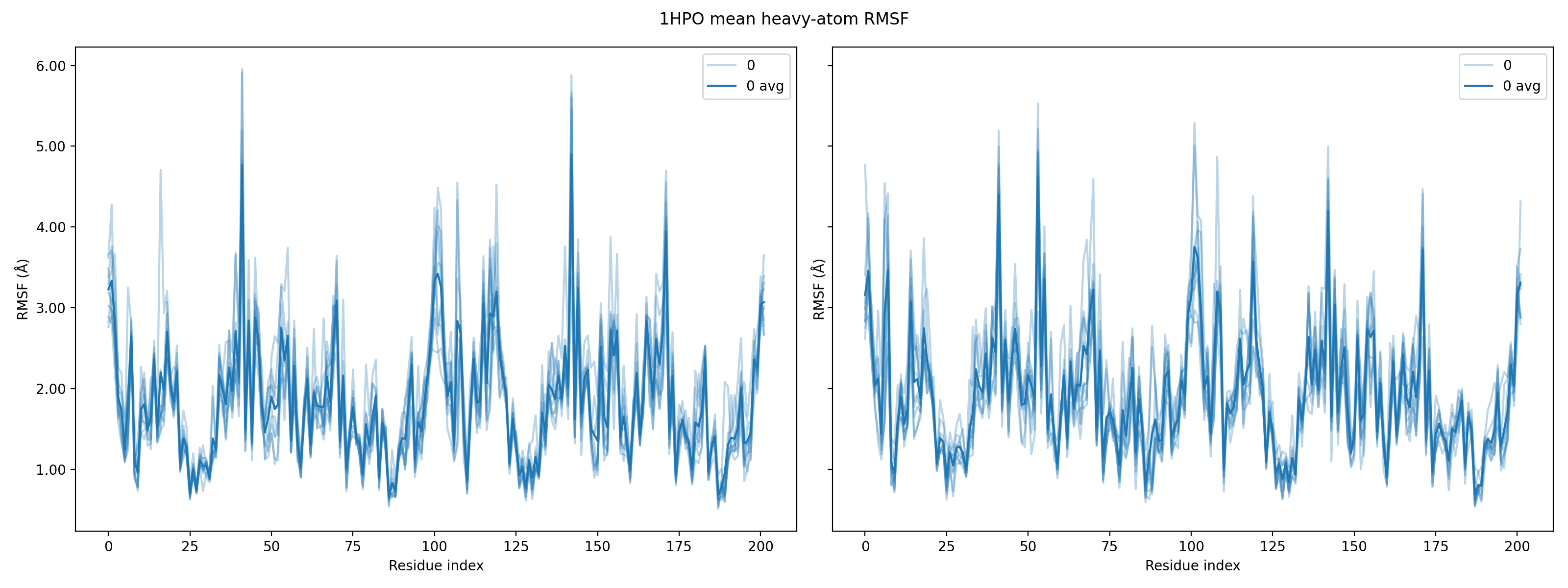}
    \caption{Residue RMSF of protein for 1HPO with the different methods (MM and NNP/MM). The independent simulations are colored in light blue and the average is in dark blue.}
\end{figure}
\begin{figure}
    \centering
    \hspace{0.05\textwidth}MM\hspace{0.35\textwidth}NNP/MM
    \includegraphics[width=0.9\textwidth]{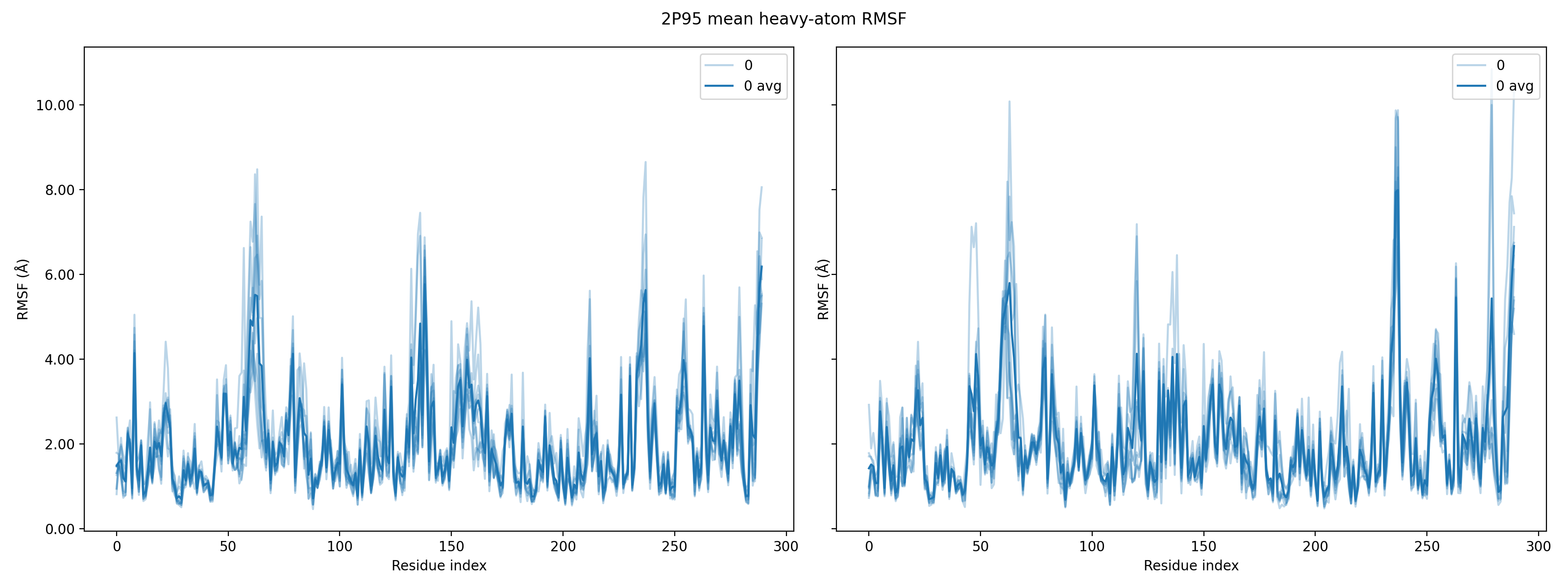}
    \caption{Residue RMSF of protein for 2P95 with the different methods (MM and NNP/MM). The independent simulations are colored in light blue and the average is in dark blue.}
\end{figure}
\begin{figure}
    \centering
    \hspace{0.05\textwidth}MM\hspace{0.35\textwidth}NNP/MM
    \includegraphics[width=0.9\textwidth]{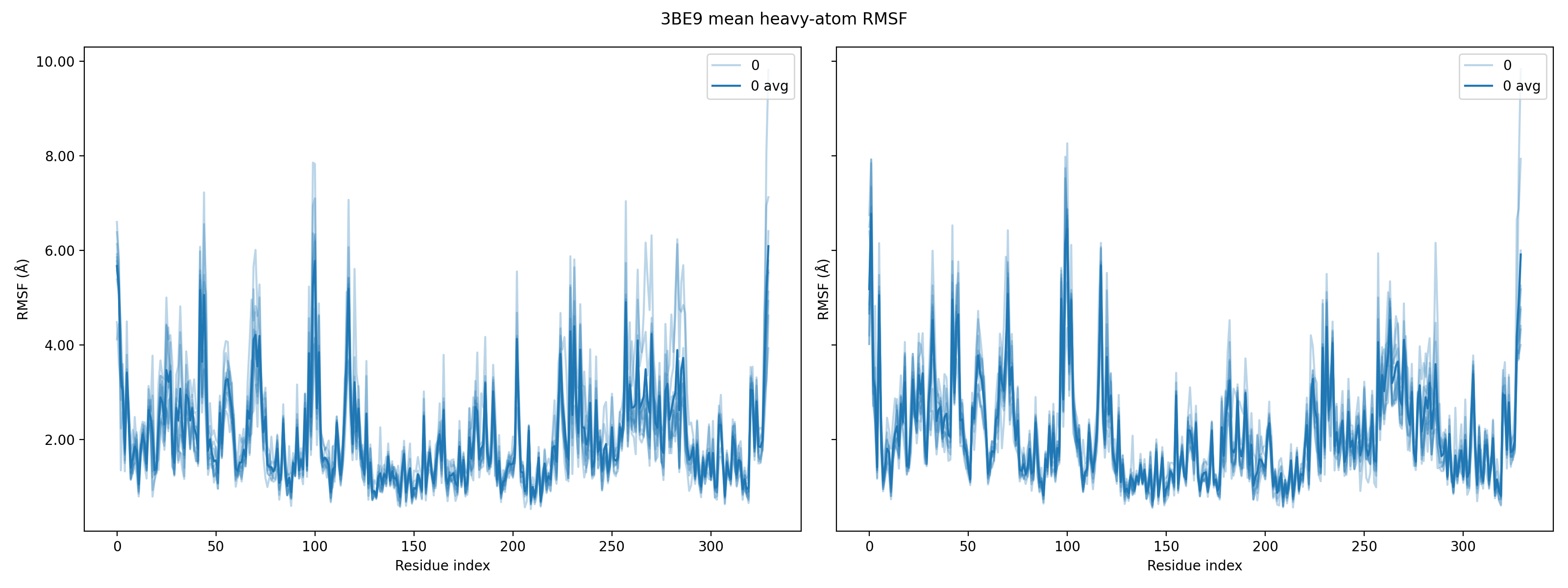}
    \caption{Residue RMSF of protein for 3BE9 with the different methods (MM and NNP/MM). The independent simulations are colored in light blue and the average is in dark blue.}
\end{figure}

\clearpage
\section{Time series of ligand RMSD}

\begin{figure}
    \centering
    \hspace{0.05\textwidth}MM\hspace{0.35\textwidth}NNP/MM
    \includegraphics[width=0.9\textwidth]{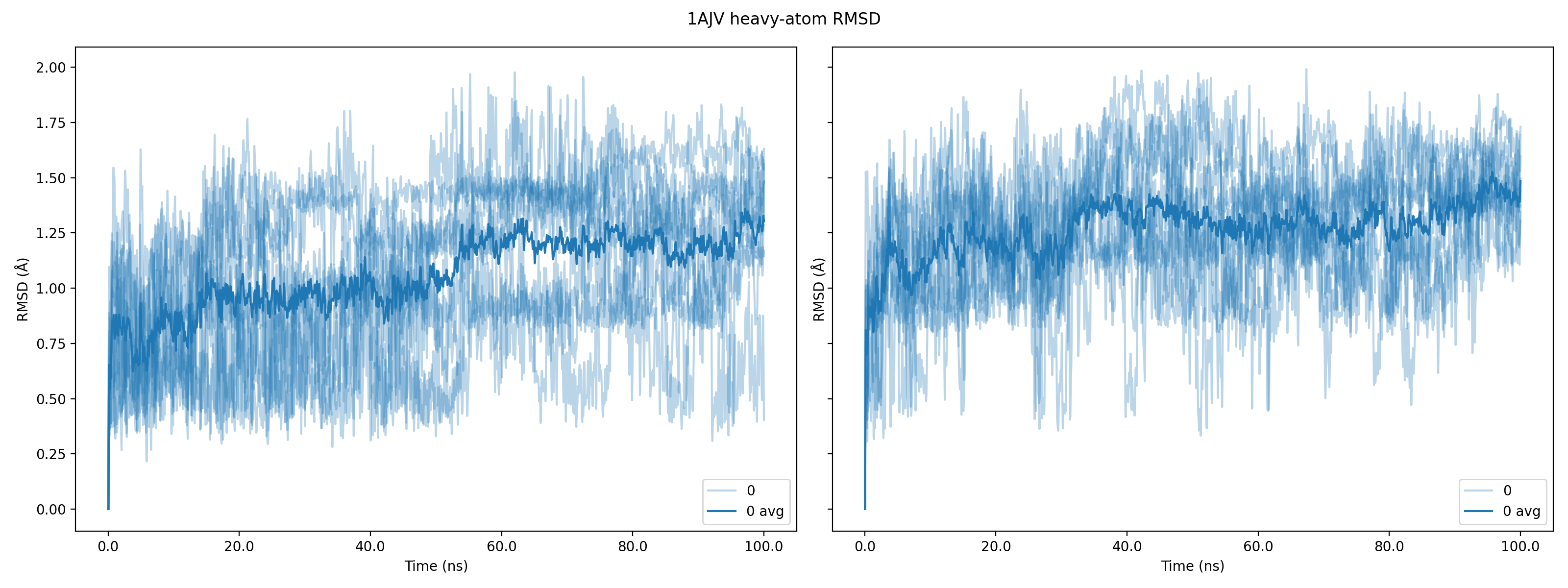}
    \caption{Time series of ligand RMSD for 1AJV with the different methods (MM and NNP/MM). The independent simulations are colored in light blue and the average is in dark blue.}
\end{figure}
\begin{figure}
    \centering
    \hspace{0.05\textwidth}MM\hspace{0.35\textwidth}NNP/MM
    \includegraphics[width=0.9\textwidth]{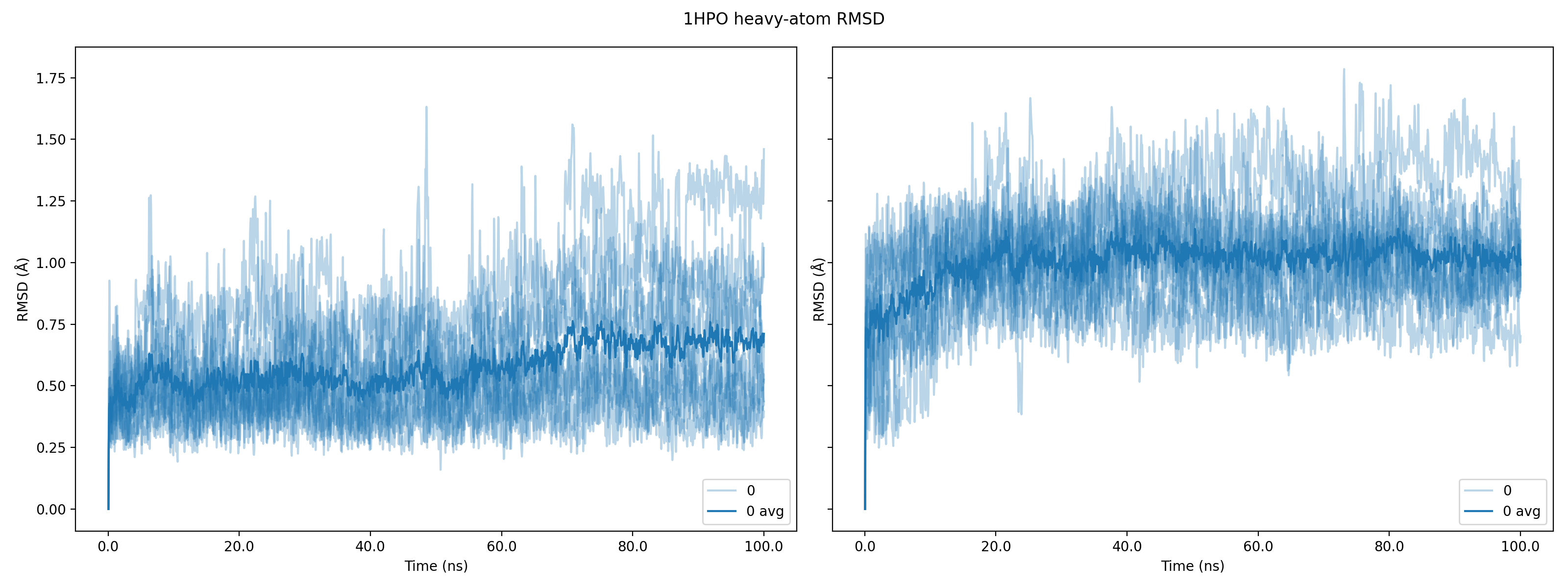}
    \caption{Time series of ligand RMSD for 1HPO with the different methods (MM and NNP/MM). The independent simulations are colored in light blue and the average is in dark blue.}
\end{figure}
\begin{figure}
    \centering
    \hspace{0.05\textwidth}MM\hspace{0.35\textwidth}NNP/MM
    \includegraphics[width=0.9\textwidth]{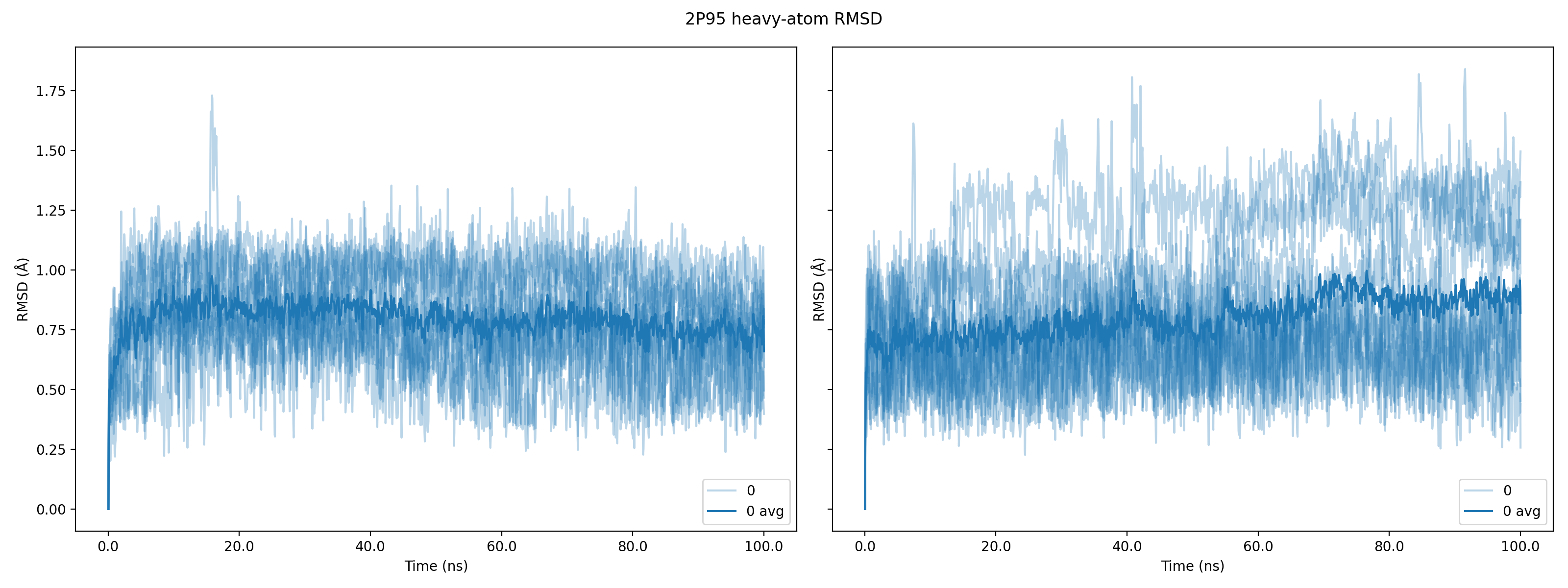}
    \caption{Time series of ligand RMSD for 2P95 with the different methods (MM and NNP/MM). The independent simulations are colored in light blue and the average is in dark blue.}
\end{figure}
\begin{figure}
    \centering
    \hspace{0.05\textwidth}MM\hspace{0.35\textwidth}NNP/MM
    \includegraphics[width=0.9\textwidth]{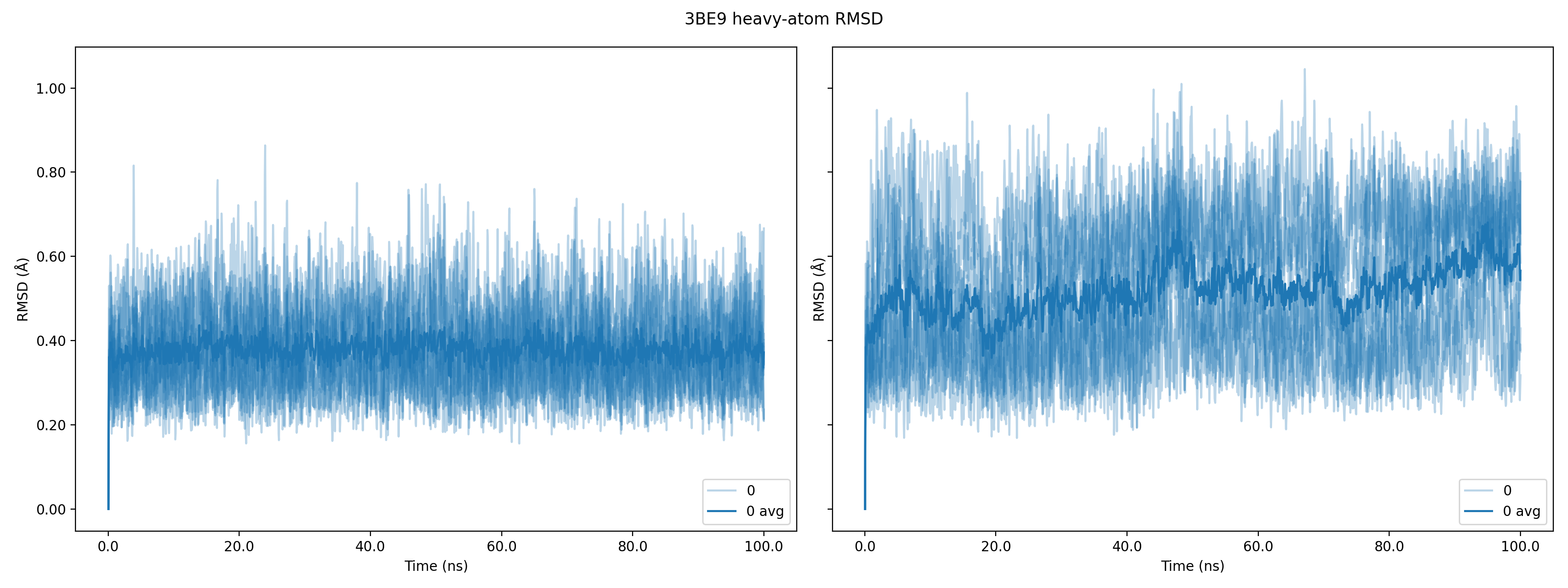}
    \caption{Time series of ligand RMSD for 3BE9 with the different methods (MM and NNP/MM). The independent simulations are colored in light blue and the average is in dark blue.}
\end{figure}

\clearpage
\section{Protein-ligand interactions}

Protein-ligand interactions have been detected following criteria:
\begin{itemize}
    \item H-bond: the donor and acceptor distance (\ch{D-A}) is <\SI{2.5}{\angstrom} and the \ch{D-H-A} angle is >\ang{120}.
    \item $\pi$-$\pi$: the distance between the centers of aromatic rings (\ch{R-R}) and the angle between their norms  (N) satisfying one of these criteria:
    \begin{itemize}
        \item \ch{R-R} is <\SI{4.4}{\angstrom} and N is <\ang{30}.
        \item \ch{R-R} is <\SI{5.5}{\angstrom} and N is >\ang{60}.
    \end{itemize}
    \item Cation-$\pi$: the distance between a positively-charged atom (P) and the center of the aromatic ring (R) is <\SI{5.0}{\angstrom} and the angle of \ch{P-R} with respect to the plane of R is >\ang{60}.
    \item $\sigma$-hole: the distance between halogen atom (H) and the center of the aromatic ring (R) is <\SI{4.0}{\angstrom} and the angle of \ch{H-R} with respect to the plane of R is >\ang{60}.
\end{itemize}
The trajectories have been sampled every \SI{50}{ps} and the probabilities computed aggregating data from all the 10~independent simulations.

\begin{table}
\caption{Probability of 1AJV protein-ligand interactions in the MM simulations. The interactions with a lower probability than 0.01 are excluded.}
\begin{threeparttable}
\begin{tabular}{ccc}
\toprule
Interaction & Residue & Probability \\
\midrule
H-bond & ASP:26  & 0.87665 \\
H-bond & ASP:26  & 0.99235 \\
H-bond & ILE:51  & 0.96835 \\
H-bond & ASH:127 & 0.26010 \\
H-bond & ASH:127 & 0.51815 \\
H-bond & ILE:152 & 0.88880 \\
\bottomrule
\end{tabular}
\end{threeparttable}
\end{table}

\begin{table}
\caption{Probability of 1AJV protein-ligand interactions in the NNP/MM simulations. The interactions with a lower probability than 0.01 are excluded.}
\begin{threeparttable}
\begin{tabular}{ccc}
\toprule
Interaction & Residue & Probability \\
\midrule
H-bond & ASP:26  & 0.99210 \\
H-bond & ASP:26  & 0.99140 \\
H-bond & ILE:51  & 0.96035 \\
H-bond & ASH:127 & 0.05240 \\
H-bond & ASH:127 & 0.11475 \\
H-bond & ILE:152 & 0.91555 \\
\bottomrule
\end{tabular}
\end{threeparttable}
\end{table}

\begin{table}
\caption{Probability of 1HPO protein-ligand interactions in the MM simulations. The interactions with a lower probability than 0.01 are excluded.}
\begin{threeparttable}
\begin{tabular}{ccc}
\toprule
Interaction & Residue & Probability \\
\midrule
Cation-$\pi$ & ARG:9   & 0.43165 \\
H-bond       & ASH:26  & 0.11160 \\
H-bond       & ILE:51  & 0.36980 \\
H-bond       & ILE:51  & 0.02740 \\
H-bond       & ASP:127 & 0.94340 \\
H-bond       & ASP:131 & 0.28660 \\
H-bond       & GLY:150 & 0.92555 \\
H-bond       & GLY:150 & 0.57730 \\
H-bond       & GLY:151 & 0.01885 \\
H-bond       & ILE:152 & 0.90360 \\
\bottomrule
\end{tabular}
\end{threeparttable}
\end{table}

\begin{table}
\caption{Probability of 1HPO protein-ligand interactions in the NNP/MM simulations. The interactions with a lower probability than 0.01 are excluded.}
\begin{threeparttable}
\begin{tabular}{ccc}
\toprule
Interaction & Residue & Probability \\
\midrule
Cation-$\pi$ & ARG:9   & 0.76190 \\
H-bond       & ASH:26  & 0.10260 \\
H-bond       & ILE:51  & 0.38165 \\
H-bond       & ILE:51  & 0.10210 \\
H-bond       & ASP:127 & 0.99580 \\
H-bond       & ASP:131 & 0.81165 \\
H-bond       & ASP:132 & 0.01970 \\
H-bond       & GLY:150 & 0.99445 \\
H-bond       & GLY:150 & 0.11820 \\
H-bond       & ILE:152 & 0.89975 \\
\bottomrule
\end{tabular}
\end{threeparttable}
\end{table}

\begin{table}
\caption{Probability of 2P95 protein-ligand interactions in the MM simulations. The interactions with a lower probability than 0.01 are excluded.}
\begin{threeparttable}
\begin{tabular}{ccc}
\toprule
Interaction & Residue & Probability \\
\midrule
$\pi$-$\pi$ & TYR:86  & 0.62565 \\
$\pi$-$\pi$ & TYR:86  & 0.10760 \\
$\pi$-$\pi$ & PHE:163 & 0.03485 \\
$\pi$-$\pi$ & PHE:163 & 0.57840 \\
$\pi$-$\pi$ & TRP:206 & 0.56830 \\
H-bond      & GLY:207 & 0.90965 \\
H-bond      & GLY:207 & 0.05580 \\
H-bond      & GLY:209 & 0.35475 \\
$\sigma$-hole & TYR:219 & 0.54095 \\
\bottomrule
\end{tabular}
\end{threeparttable}
\end{table}

\begin{table}
\caption{Probability of 2P95 protein-ligand interactions in the NNP/MM simulations. The interactions with a lower probability than 0.01 are excluded.}
\begin{threeparttable}
\begin{tabular}{ccc}
\toprule
Interaction & Residue & Probability \\
\midrule
$\pi$-$\pi$ & TYR:86  & 0.63505 \\
$\pi$-$\pi$ & TYR:86  & 0.08905 \\
$\pi$-$\pi$ & PHE:163 & 0.06140 \\
$\pi$-$\pi$ & PHE:163 & 0.59525 \\
$\pi$-$\pi$ & TRP:206 & 0.01760 \\
$\pi$-$\pi$ & TRP:206 & 0.64465 \\
H-bond      & GLY:207 & 0.89720 \\
H-bond      & GLY:207 & 0.07225 \\
H-bond      & GLY:209 & 0.54680 \\
$\sigma$-hole & TYR:219 & 0.38280 \\
\bottomrule
\end{tabular}
\end{threeparttable}
\end{table}

\begin{table}
\caption{Probability of 3BE9 protein-ligand interactions in the MM simulations. The interactions with a lower probability than 0.01 are excluded.}
\begin{threeparttable}
\begin{tabular}{ccc}
\toprule
Interaction & Residue & Probability \\
\midrule
H-bond      & LYS:64  & 0.68070 \\
H-bond      & VAL:112 & 0.57070 \\
H-bond      & VAL:112 & 0.38385 \\
$\pi$-$\pi$ & HID:156 & 0.24210 \\
H-bond      & ASP:171 & 0.26770 \\
H-bond      & ASP:171 & 0.10400 \\
\bottomrule
\end{tabular}
\end{threeparttable}
\end{table}

\begin{table}
\caption{Probability of 3BE9 protein-ligand interactions in the NNP/MM simulations. The interactions with a lower probability than 0.01 are excluded.}
\begin{threeparttable}
\begin{tabular}{ccc}
\toprule
Interaction & Residue & Probability \\
\midrule
H-bond      & VAL:41  & 0.04610 \\
H-bond      & LYS:64  & 0.78265 \\
H-bond      & VAL:112 & 0.65315 \\
H-bond      & VAL:112 & 0.98575 \\
$\pi$-$\pi$ & HID:156 & 0.24225 \\
H-bond      & ASP:171 & 0.10545 \\
H-bond      & ASP:171 & 0.03135 \\
\bottomrule
\end{tabular}
\end{threeparttable}
\end{table}

\clearpage
\bibliography{references}